\documentclass{tmp}

\def\ie{{\it i.e.},~}
\def\eg{{\it e.g.},~}
\def\4he{$^4$He}
\def\3he{$^3$He}
\def\7li{$^7$Li}
\def\Yp{Y$_{\rm P}$~}
\def\yd{$y_{\rm D}$~}
\def\yli{$y_{\rm Li}$~}
\def\xie{$\xi_{e}$~}
\def\hii{H\thinspace{$\scriptstyle{\rm II}$}~}
\def\hi{H\thinspace{$\scriptstyle{\rm I}$}~}  
\def\di{D\thinspace{$\scriptstyle{\rm I}$}~}  

\def\Nnu{N$_{\nu}$~}
\newcommand{\epm}{\ensuremath{e^{\pm}\;}}
\newcommand{\be}{\begin{equation}}
\newcommand{\ee}{\end{equation}}
\newcommand{\Deln}{\ensuremath{\Delta N_\nu\;}}
\def\Nnu{N$_{\nu}$~}

\newcommand\la{\lower0.6ex\vbox{\hbox{\ensuremath{\buildrel{\textstyle<}\over{\sim}\ }}}}
\newcommand\ga{\lower0.6ex\vbox{\hbox{\ensuremath{\buildrel{\textstyle>}\over{\sim}\ }}}}

\begin{document}

\markboth{Gary Steigman}
{PRIMORDIAL NUCLEOSYNTHESIS}

\catchline{}{}{}{}{}
%
%%%%%%%%%%%%%%%%%%%%%%%%%%%%%%%%%%%%%%%%%%%%%%%%%%%%%%%%%%%%%%%%%%%%

\title{PRIMORDIAL NUCLEOSYNTHESIS:\\ 
SUCCESSES AND CHALLENGES}

\author{\footnotesize GARY STEIGMAN}

\address{Physics Department, The Ohio State University, 
191 West Woodruff Avenue\\
Columbus, Ohio 43210, USA}

\maketitle

\begin{history}
\received{15 November 2005}
%\revised{(revised date)}
\end{history}

\begin{abstract}
Primordial nucleosynthesis provides a probe of the Universe during 
its early evolution.  Given the progress exploring the constituents, 
structure, and recent evolution of the Universe, it is timely to 
review the status of Big Bang Nucleosynthesis (BBN) and to confront 
its predictions, and the constraints which emerge from them, with 
those derived from independent observations of the Universe at much 
later epochs in its evolution.  Following an overview of the key 
physics controlling element synthesis in the early Universe, the 
predictions of the standard models of cosmology and particle physics 
are presented, along with those from some non-standard models.  
The observational data used to infer the primordial abundances are 
described, with an emphasis on the distinction between {\it precision} 
and {\it accuracy}.  These relic abundances are compared with predictions, 
testing the internal consistency of BBN and enabling a comparison of 
the BBN constraints with those derived from the WMAP Cosmic Background 
Radiation data.  Emerging from these comparisons is a successful 
standard model along with constraints on (or hints of) physics 
beyond the standard models of particle physics and of cosmology.

\end{abstract}

\section{Introduction}	

As the Universe evolved from its early, hot, dense beginnings (the 
``Big Bang'') to its present, cold, dilute state, it passed through 
a brief epoch when the temperature (average thermal energy) and 
density of its nucleon component were such that nuclear reactions 
building complex nuclei could occur.  Because the nucleon content 
of the Universe is small (in a sense to be described below) and 
because the Universe evolved through this epoch very rapidly, only 
the lightest nuclides (D, \3he, \4he, and \7li) could be synthesized 
in astrophysically interesting abundances.  The relic abundances 
of these nuclides provide probes of conditions and contents of 
the Universe at a very early epoch in its evolution (the first 
few minutes) otherwise hidden from our view.  The standard model
of Cosmology subsumes the standard model of particle physics (\eg
three families of very light, left-handed neutrinos along with their
right-handed antineutrinos) and uses General Relativity (\eg the
Friedman equation) to track the time-evolution of the universal
expansion rate and its matter and radiation contents.  While nuclear 
reactions among the nucleons are always occurring in the early Universe, 
Big Bang Nucleosynthesis (BBN) begins in earnest when the Universe 
is a few minutes old and it ends less than a half hour later when 
nuclear reactions are quenched by low temperatures and densities.  
The BBN abundances depend on the conditions (temperature, nucleon 
density, expansion rate, neutrino content and neutrino-antineutrino 
asymmetry, etc.) at those times and are largely independent of the 
detailed processes which established them.  As a consequence, BBN 
can test and constrain the parameters of the standard model (SBBN), 
as well as probe any non-standard physics/cosmology which changes 
those conditions.

The relic abundances of the light nuclides synthesized in BBN depend on 
the competition between the nucleon density-dependent nuclear reaction 
rates and the universal expansion rate.  In addition, while all 
primordial abundances depend to some degree on the initial (when BBN 
begins) ratio of neutrons to protons, the \4he abundance is largely 
fixed by this ratio, which is determined by the competition between 
the weak interaction rates and the universal expansion rate, along 
with the magnitude of any $\nu_{e} - \bar\nu_{e}$ asymmetry.\footnote{A 
lepton asymmetry much larger than the baryon asymmetry (which is very 
small; see \S1.1 below) would have to reside in the neutrinos since 
charge neutrality ensures that the electron-positron asymmetry is 
comparable to the baryon asymmetry.}  To summarize, in its simplest 
version BBN depends on three unknown parameters: the baryon asymmetry; 
the lepton asymmetry; the universal expansion rate.  These parameters 
are quantified next.

\subsection{Baryon Asymmetry -- Nucleon Abundance}

In the very early universe baryon-antibaryon pairs (quark-antiquark pairs) 
were as abundant as radiation (\eg photons).  As the Universe expanded and 
cooled, the pairs annihilated, leaving behind any baryon {\bf excess} 
established during the earlier evolution of the Universe\cite{anti}.  
Subsequently, the number of baryons in a comoving volume of the Universe 
is preserved.  After \epm pairs annihilate, when the temperature (in energy 
units) drops below the electron mass, the number of Cosmic Background 
Radiation (CBR) photons in a comoving volume is also preserved.  As a 
result, it is useful (and conventional) to measure the universal baryon 
asymmetry by comparing the number of ({\it excess}) baryons to the number 
of photons in a comoving volume (post-\epm annihilation).  This ratio 
defines the baryon abundance parameter $\eta_{\rm B}$,
\be
\eta_{\rm B} \equiv {n_{\rm B} - n_{\bar{\rm B}} \over n_{\gamma}}.
\ee
As will be seen from BBN, and as is confirmed by a variety of 
independent (non-BBN), astrophysical and cosmological data, 
$\eta_{\rm B}$ is very small.  As a result, it is convenient to 
introduce $\eta_{10} \equiv 10^{10}\eta_{\rm B}$ and to use it 
as one of the adjustable parameters for BBN.  An equivalent measure 
of the baryon density is provided by the baryon density parameter, 
$\Omega_{\rm B}$, the ratio (at present) of the baryon mass density 
to the critical density.  In terms of the present value of the Hubble 
parameter (see \S 1.2 below), $H_{0} \equiv 100h$~km~s$^{-1}$~Mpc$^{-1}$, 
these two measures are related by 
\be
\eta_{10} \equiv 10^{10}(n_{\rm B}/n_{\gamma})_{0} = 274\Omega_{\rm B}h^{2}.  
\ee
Note that the subscript 0 refers to the present epoch (redshift $z = 0$).

From a variety of non-BBN cosmological observations whose accuracy 
is dominated by the very precise CBR temperature fluctuation data 
from WMAP\cite{sperg}, the baryon abundance parameter is limited to 
a narrow range centered near $\eta_{10} \approx 6$.  As a result, 
while the behavior of the BBN-predicted relic abundances will be 
described qualitatively as functions of $\eta_{\rm B}$, for 
quantitative comparisons the results presented here will focus on 
the limited interval $4 \le \eta_{10} \le 8$.  As will be seen below 
(\S2.2), over this range there are very simple, yet accurate, analytic 
fits to the BBN-predicted primordial abundances.

\subsection{The Expansion Rate At BBN}

For the standard model of cosmology, the Friedman equation relates
the expansion rate, quantified by the Hubble parameter ($H$), to
the matter-radiation content of the Universe.
\be
H^{2} = {8\pi \over 3}G_{\rm N}\rho_{\rm TOT},
\label{friedman}
\ee
where $G_{\rm N}$ is Newton's gravitational constant.  During the 
early evolution of the Universe the total density, $\rho_{\rm TOT}$, 
is dominated by ``radiation'' (\ie by the contributions from massless 
and/or extremely relativistic particles).  During radiation dominated 
epochs (RD), the age of the Universe ($t$) and the Hubble parameter 
are simply related by $(Ht)_{\rm RD} = 1/2$.

Prior to BBN, at a temperature of a few MeV, the standard model of 
particle physics determines that the relativistic particle content 
consists of photons, \epm pairs and three flavors of left-handed 
(\ie one helicity state) neutrinos (along with their right-handed, 
antineutrinos; N$_{\nu} = 3$).  With all chemical potentials set 
to zero (very small lepton asymmetry) the energy density of these 
constituents in thermal equilibrium is  
\be 
\rho_{\rm TOT} = \rho_{\gamma} + \rho_{e} + 3\rho_{\nu} = {43 \over 8}\rho_{\gamma}, 
\label{rho0} 
\ee 
where $\rho_{\gamma}$ is the energy density in the CBR photons (which 
have redshifted to become the CBR photons observed today at a temperature 
of 2.7K).  In this case (SBBN: N$_{\nu} = 3$), the time-temperature 
relation derived from the Friedman equation is,  
\be 
{\rm Pre-\epm annihilation}:~~t~T_{\gamma}^{2} = 0.738~{\rm MeV^{2}~s}. 
\label{ttpre} 
\ee 
 
In SBBN it is usually assumed that the neutrinos are fully decoupled 
prior to \epm annihilation; if so, they don't share in the energy 
transferred from the annihilating \epm pairs to the CBR photons.  In 
this very good approximation, the photons are hotter than the neutrinos 
in the post-\epm annihilation universe by a factor $T_{\gamma}/T_{\nu} = 
(11/4)^{1/3}$, and the total energy density is 
\be 
\rho_{\rm TOT} = \rho_{\gamma} + 3\rho_{\nu} = 1.68\rho_{\gamma}, 
\ee 
corresponding to a modified time-temperature relation, 
\be 
{\rm Post-\epm annihilation}:~~t~T_{\gamma}^{2} = 1.32~{\rm MeV^{2}~s}. 
\label{ttpost} 
\ee 
 
Quite generally, new physics beyond the standard models of cosmology 
or particle physics could lead to a non-standard, early Universe 
expansion rate ($H'$), whose ratio to the standard rate ($H$) may be 
parameterized by an expansion rate factor $S$, 
\be 
H \rightarrow H' \equiv SH. 
\label{S} 
\ee 
A non-standard expansion rate might originate from modifications to 
the 3+1 dimensional Friedman equation as in some higher dimensional 
models\cite{rs}, or from a change in the strength of gravity\cite{KS2003}.
Different gravitational couplings for fermions and bosons\cite{BS2004} 
would have similar effects.  Alternatively, changing the particle 
population in early Universe will modify the energy density -- temperature
relation, also leading, through eq.~\ref{friedman}, to $S \neq 1$.  While 
these different mechanisms for implementing a non-standard expansion  
rate are not necessarily equivalent, specific models generally lead 
to specific predictions for $S$.  

Consider, for example, the case of a non-standard energy density.   
\be 
\rho_{\rm R} \rightarrow 
\rho_{\rm R}' \equiv S^{2}\rho_{\rm R}, 
\label{rho'} 
\ee 
where $\rho_{\rm R}' = \rho_{\rm R} + \rho_{X}$ and $X$ identifies 
the non-standard component.  With the restriction that the $X$ 
are relativistic, this extra component, non-interacting at \epm 
annihilation, behaves as would an additional neutrino flavor.  
It must be emphasized that $X$ is {\bf not} restricted to additional 
flavors of active or sterile neutrinos.  In this class of models 
$S$ is constant prior to \epm annihilation and it is convenient 
(and conventional) to account for the extra contribution to the 
standard-model energy density by normalizing it to that of an 
``equivalent" neutrino flavor\cite{ssg}, so that  
\be 
\rho_{X} \equiv \Delta N_{\nu}\rho_{\nu} = 
{7 \over 8}\Delta N_{\nu}\rho_{\gamma}. 
\label{deln} 
\ee 
For this case, 
\be 
S \equiv S_{pre} = \bigg(1 + {7 \over 43}\Delta N_{\nu}\bigg)^{1/2}. 
\label{sdeltannu} 
\ee 

In another class of non-standard models the early Universe is 
heated by the decay of a massive particle, produced earlier in 
the evolution\cite{lowreheat}.  If the Universe is heated to a 
temperature which is too low to (re)populate a thermal spectrum 
of the standard neutrinos ($T_{\rm RH}~\la 7$~MeV), the effective 
number of neutrino flavors contributing to the total energy 
density is $< 3$, resulting in \Deln $< 0$ and $S < 1$.

Since the expansion rate is more fundamental than is $\Delta N_{\nu}$, 
BBN for models with non-standard expansion rates will be parameterized 
using $S$ (but, for comparison, the corresponding value of \Deln from 
eq.~\ref{sdeltannu} will often be given for comparison).  The simple, 
analytic fits to BBN presented below (\S2.2) are quite accurate for 
$0.85 \le S \le 1.15$, corresponding to $-1.7 ~\la \Delta$N$_{\nu} ~\la 
2.0$

\subsection{Neutrino Asymmetry}

The baryon asymmetry of the Universe, quantified by $\eta_{\rm B}$, is 
very small.  If, as expected in the currently most popular particle 
physics models, the universal lepton and baryon numbers are comparable, 
then any asymmetry between neutrinos and antineutrinos (``neutrino 
degeneracy'') will be far too small to have a noticeable effect on 
BBN.  However, it is possible that the baryon and lepton asymmetries 
are disconnected and that the lepton (neutrino) asymmetry could be large 
enough to perturb the SBBN predictions.  In analogy with $\eta_{\rm B}$ 
which quantifies the baryon asymmetry, the lepton (neutrino) asymmetry, 
$L = L_{\nu} \equiv \Sigma_{\alpha} L_{\nu_{\alpha}}$, may be quantified 
by the neutrino chemical potentials $\mu_{\nu_{\alpha}}$ ($\alpha \equiv 
e, \mu, \tau$) or, by the degeneracy parameters, the ratios of the 
neutral lepton chemical potentials to the temperature (in energy units) 
$\xi_{\nu_{\alpha}} \equiv \mu_{\nu_{\alpha}}/kT$, where
\be 
L_{\nu_{\alpha}} \equiv \bigg({n_{\nu_{\alpha}}-n_{\bar\nu_{\alpha}} 
\over n_{\gamma}}\bigg)= {\pi^2 \over 12 \zeta(3)}\bigg(
{T_{\nu_{\alpha}} \over T_{\gamma}}\bigg)^{3}
\bigg(\xi_{\nu_{\alpha}}+{\xi_{\nu_{\alpha}}^3 \over \pi^2}\bigg)\,. 
\label{lasym}
\ee 
Prior to \epm annihilation, $T_{\nu} = T_{\gamma}$, while post-\epm 
annihilation $(T_{\nu}/ T_{\gamma})^{3} = 4/11$.  Although in principle 
the asymmetry among the different neutrino flavors may be different, 
mixing among the three active neutrinos ($\nu_{e}, \nu_{\mu}, \nu_{\tau}$) 
ensures that at BBN, $L_{e} \approx L_{\mu} \approx L_{\tau}$ ($\xi_{e} 
\approx \xi_{\mu} \approx \xi_{\tau}$)\cite{equal}. If $L_{\nu}$ is 
measured {\bf post}-\epm annihilation, as is $\eta_{\rm B}$, then for 
$\xi_{\nu} \ll 1$, $L_{\nu} \approx 3L_{\nu_{e}}$ and, for $\xi \equiv 
\xi_{e} \ll 1$, $L_{\nu} \approx 0.75\xi$.

Although any neutrino degeneracy ($\xi_{\nu_{\alpha}} < 0$ as well as 
$> 0$) {\it increases} the energy density in the relativistic neutrinos,
resulting in an {\it effective} \Deln$ \neq 0$ (see eq.~\ref{deln}), the 
range of $|\xi|$ of interest to BBN is limited to sufficiently small 
values that the increase in $S$ due to a non-zero $\xi$ is negligible. 
However, a small asymmetry between {\it electron} type neutrinos and 
antineutrinos ($\xi_{e} ~\ga 10^{-2}$; $L ~\ga 0.007$), while large 
compared to the baryon asymmetry, can have a significant impact on 
BBN since the $\nu_{e}$ affect the interconversion of neutrons to 
protons.  A non-zero $\xi_{e}$ results in different (compared to 
SBBN) numbers of $\nu_{e}$ and $\bar\nu_{e}$, altering the n/p ratio 
at BBN, thereby changing the yields (compared to SBBN) of the light 
nuclides.   
 
Of the light, relic nuclei, the {\bf neutron limited} \4he abundance is 
most sensitive to a non-zero $\xi_{e}$; \4he is a good ``leptometer''.  
In concert with the abundances of D, \3he, and \7li, which are good 
baryometers, the \4he abundance provides a test of the consistency 
of the standard model along with constraints on non-standard models.  
The analytic fits presented below (\S2.2) are reasonably accurate 
for $\xi_{e}$ in the range, $-0.1~\la \xi_{e}~\la 0.1$, corresponding 
to a total lepton number limited to $|L|~\la 0.07$.  While this may 
seem small, recall that a similar measure of the baryon asymmetry is 
orders of magnitude smaller: $\eta_{\rm B} \approx 6\times 10^{-10}$.

\section{An Overview of Primordial Nucleosynthesis}

The early ($\ga 1$~ms), hot, dense Universe is filled with radiation 
($\gamma$s, \epm pairs, $\nu$s of all flavors), along with dynamically 
and numerically insignificant amounts of baryons (nucleons) and dark 
matter particles.  Nuclear and weak interactions are occurring among
the neutrons, protons, \epm, and $\nu$s (\eg $n+p \longleftrightarrow 
D+\gamma$; $p+e^{-} \longleftrightarrow n+\nu_{e}$) at rates fast 
compared to the universal expansion rate.  At such high temperatures 
($T~\ga 3$~MeV), in an environment where the nucleon to photon ratio 
is very small ($\eta_{10} \approx 3 - 10$), the abundances of complex 
nuclei (D, \3he,\4he, \7li) are tiny in comparison to those of the 
free nucleons (neutrons and protons).  At the same time, the 
charged-current weak interactions are regulating the neutron to proton 
ratio, initially keeping it close to its equilibrium value
\be
(n/p)_{eq} = e^{-\Delta m/T},
\label{n/peq}
\ee
where $\Delta m$ is the neutron -- proton mass (energy) difference.  
In this context it is worth noting that if there is an {\it asymmetry} 
between the numbers of $\nu_{e}$ and $\bar\nu_{e}$ the equilibrium 
neutron-to-proton ratio is modified to $(n/p)_{eq} = \exp(-\Delta 
m/T - \mu_{e}/T) = e^{-\xi_{e}}(n/p)^{0}_{eq}$.

As the Universe expands and cools, the lighter protons are favored 
over the heavier neutrons and the neutron-to-proton ratio decreases, 
tracking the equilibrium form in eq.~\ref{n/peq}.  But, as the temperature 
decreases below $T \sim 0.8$~MeV, when the Universe is $\sim 1$~second 
old, the weak interactions are too slow to maintain equilibrium and 
the neutron-to-proton ratio, while continuing to fall, deviates from 
({\it exceeds}) the equilibrium value.  Since the $n/p$ ratio depends 
on the {\it competition} between the weak interaction rates and the 
early-Universe expansion rate (as well as on a possible neutrino 
asymmetry), deviations from the standard model (\eg $\rho_{\rm R} 
\rightarrow \rho_{\rm R} + \rho_{X}$ or $\xi_{e} \neq 0$) will change 
the relative numbers of neutrons and protons available for building 
the complex nuclides.    

As noted above, while neutrons and protons are interconverting, they 
are also colliding among themselves creating complex nuclides, \eg 
deuterons.  However, at early times, when the density and average 
energy of the CBR photons are very high, the newly formed deuterons 
find themselves bathed in a background of high-energy gamma rays 
capable of photodissociating them.  Since there are more than a 
billion CBR photons for every nucleon in the Universe, the deuteron 
is photodissociated before it can capture a neutron (or a proton, or 
another deuteron) to build the heavier nuclides.  This {\it bottleneck} 
to BBN persists until the temperature drops sufficiently below the 
binding energy of the deuteron, when there are too few photons energetic 
enough to photodissociate them before they capture nucleons, launching 
BBN.  This transition (smooth, but rapid) occurs after \epm annihilation, 
when the Universe is a few minutes old and the temperature has dropped 
below $\sim 80$~keV.   

Once BBN begins in earnest, neutrons and protons quickly combine to build  
D, $^3$H, \3he, and \4he.  Since there are no stable mass-5 nuclides, a
new bottleneck appears at \4he.  Nuclear reactions quickly incorporate
all available neutrons into \4he, the most strongly bound of the light
nuclides.  Jumping the gap at mass-5 requires Coulomb suppressed reactions 
of \4he with D, or $^3$H, or \3he, guaranteeing that the abundances of 
the heavier nuclides are severely depressed below that of \4he (and 
even of D and \3he), and that the \4he abundance is determined by the 
neutron abundance when BBN begins.  The few reactions that manage to 
bridge the mass-5 gap lead mainly to mass-7 (\7li or, to $^7$Be which, 
later, when the Universe has cooled further, will capture an electron 
and decay to \7li); for the range of $\eta_{\rm B}$ of interest, the 
BBN-predicted abundance of $^6$Li is more than 3 orders of magnitude 
below that of the more tightly bound \7li.  Finally, there is another 
gap at mass-8, ensuring that there is no astrophysically significant 
production of heavier nuclides.   

\begin{figure}
\centerline{\psfig{file=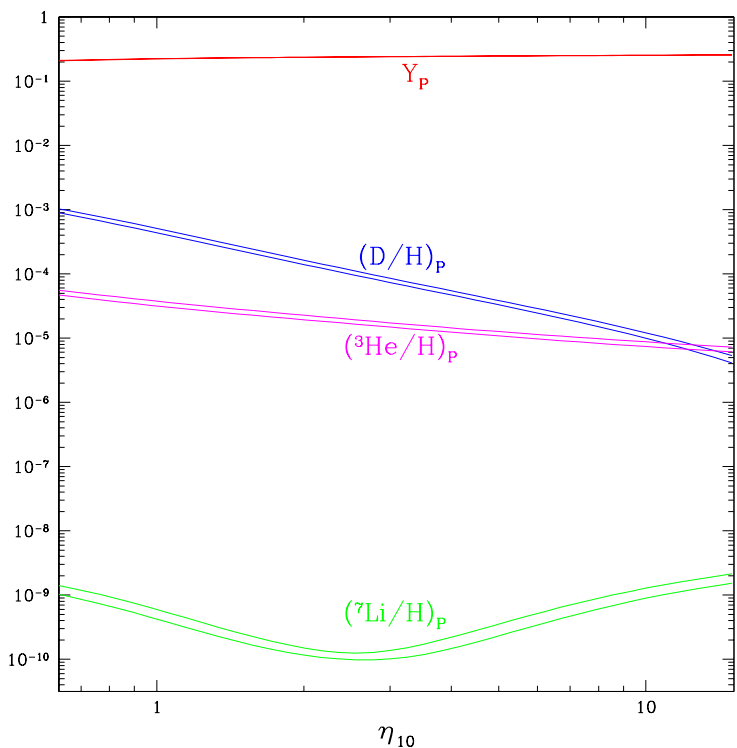,width=10.7cm}}%12cm}}
\vspace*{0pt}%\vspace*{8pt}
\caption{The SBBN-predicted primordial abundances of D, \3he, \7li (relative 
to hydrogen by number), and the \4he mass fraction (Y$_{\rm P}$), as 
functions of the baryon abundance parameter $\eta_{10}$.  The widths of 
the bands reflect the uncertainties in the nuclear and weak interaction 
rates.}
\label{fig:schrplot}
\end{figure}
   
The primordial nuclear reactor is short-lived.  As the temperature drops
below $T~\la 30$~keV, when the Universe is $\sim 20$~minutes old, Coulomb 
barriers abruptly suppress all nuclear reactions.  Afterwards, until
the first stars form, no pre-existing, primordial nuclides are destroyed 
(except for those like $^{3}$H and $^{7}$Be that are unstable and decay) 
and no new nuclides are created.  In $\sim 1000$ seconds BBN has run its 
course. 

With this as background, the trends of the SBBN-predicted primordial 
abundances of the light nuclides with baryon abundance shown in Figure 
\ref{fig:schrplot} can be understood.  The reactions burning D and \3he
(along with $^{3}$H) to \4he are very fast (compared to the universal
expansion rate) once the deuterium bottleneck is breached, ensuring
that almost all neutrons present at that time are incorporated into
\4he.  As a result, since \4he production is not {\it rate limited},
its primordial abundance is very insensitive (only logarithmically) to
the baryon abundance.  The very slight increase in \Yp with increasing
$\eta_{\rm B}$ reflects the fact that for a higher baryon abundance BBN 
begins slightly earlier, when slightly more neutrons are available.  The 
thickness of the \Yp curve in Fig.~\ref{fig:schrplot} reflects the very
small uncertainty in the BBN prediction; the uncertainty in \Yp ($\sim 
$~0.2\%; $\sigma_{\rm Y} \sim 0.0005$) is dominated by the very small 
error in the weak interaction rates which are normalized by the neutron 
lifetime ($\tau_{n} = 885.7\pm 0.8$~s).  The differences among the \Yp 
predictions from independent BBN codes are typically no larger than
$\Delta$Y$_{\rm P} \sim$~0.0002.

Nuclear reactions burn D, $^{3}$H, and \3he to \4he, the most tightly 
bound of the light nuclides, at a rate which increases with increasing 
nucleon density, accounting for the decrease in the abundances of D and 
\3he (the latter receives a contribution from the $\beta$-decay of $^{3}$H)
with higher values of $\eta_{\rm B}$.   The behavior of \7li is more 
interesting, reflecting two pathways to mass-7.  At the relatively low 
values of $\eta_{10} ~\la 3$, mass-7 is largely synthesized as \7li by 
$^3$H($\alpha$,$\gamma$)\7li reactions.  \7li is easily destroyed in 
collisions with protons.  So, for low nucleon abundance, as $\eta_{\rm B}$ 
increases, destruction is faster than production and \7li/H {\it decreases}.  
In contrast, at relatively high values of $\eta_{10} ~\ga 3$, mass-7 is 
largely synthesized as $^7$Be via \3he($\alpha$,$\gamma$)$^7$Be reactions.  
$^7$Be is more tightly bound than \7li and, therefore, harder to destroy.  
As $\eta_{\rm B}$ increases at high nucleon abundance, the primordial 
abundance of $^7$Be {\it increases}.  Later in the evolution of the Universe, 
when it is cooler and neutral atoms begin to form, $^7$Be captures an 
electron and $\beta$-decays to \7li.  These two paths to mass-7 account 
for the valley shape of the \7li abundance curve in Fig.~\ref{fig:schrplot}.   
  
Not shown on Figure~\ref{fig:schrplot} are the BBN-predicted relic abundances 
of $^6$Li, $^9$Be, $^{10}$B, and $^{11}$B.  Their production is suppressed by
the gap at mass-8.  For the same range in $\eta_{\rm B}$, all of them lie 
offscale, in the range $10^{-20} - 10^{-13}$.  

For SBBN the relic abundances of the light nuclides depend on only one 
free parameter, the nucleon abundance parameter $\eta_{\rm B}$.  As Figure 
1 reveals, for the ``interesting" range (see below) of $4~\la \eta_{10}~\la 
8$, the \4he mass fraction is expected to be Y$_{\rm P} \approx 0.25$, with 
negligible dependence on $\eta_{\rm B}$ while D/H and \3he/H decrease from 
$\approx 10^{-4}$ to $\approx 10^{-5}$, and \7li/H increases from $\approx 
10^{-10}$ to $\approx 10^{-9}$.  The light nuclide relic abundances span 
some nine orders of magnitude, yet if SBBN is correct, one choice of 
$\eta_{\rm B}$ (within the errors) should yield predictions consistent
with observations.  Before confronting the theory with data, it is useful
to consider a few generic examples of BBN in the presence of nonstandard
physics and/or cosmology.

\subsection{Nonstandard BBN}  
  
The variety of modifications to the standard models of particle physics
and of cosmology is very broad, limited only by the creativity of theorists.
Many nonstandard models introduce several, new, free parameters in addition 
to the baryon abundance parameter $\eta_{\rm B}$.  Since there are only 
four nuclides whose relic abundance is large enough to be astrophysically 
interesting and, as will be explained below in more detail, only three for 
which data directly relating to their primordial abundances exist at present 
(D, \4he, \7li), nonstandard models with two or more additional parameters 
may well be unconstrained by BBN.  Furthermore, as discussed in the 
Introduction (see \S1.2 and \S1.3), there already exist two additional 
parameters with claims to relevance: the expansion rate parameter $S$ (or, 
$\Delta N_{\nu}$; see eqs.~\ref{S},\ref{sdeltannu}) and the lepton asymmetry 
parameter $L$ (or, $\xi$; see eq.~\ref{lasym}).

The primordial abundance of \4he depends sensitively on the pre- and the 
post-\epm annihilation early universe expansion rate (the Hubble parameter 
$H$) and on the magnitude of a $\nu_{e} - \bar\nu_{e}$ asymmetry because 
each will affect the n/p ratio at BBN (see, \eg Steigman, Schramm \& Gunn 
1977 (SSG)\cite{ssg}; for recent results see Kneller \& Steigman 2004 
(KS)\cite{ks}).  A faster expansion ($S > 1$; \Deln $> 0$) leaves less time 
for neutrons to convert into protons and the higher neutron abundance 
results in increased production of \4he.  For small changes at fixed 
$\eta_{\rm B}$, $\Delta$Y$_{\rm P} \approx 0.16(S-1) \approx 0.013\Delta 
N_{\nu}$ (KS).  Although the relic abundances of D and \3he do depend 
on the competition between the nuclear reaction rates and the post-\epm 
annihilation expansion rate (faster expansion $\Rightarrow$ less D and \3he
destruction $\Rightarrow$ more D and \3he), they are much less sensitive to 
relatively small deviations from $S = 1$ (\Deln = 0)\cite{ks}.  For mass-7 
the effect of a nonstandard expansion rate is different at low and high 
values of $\eta_{\rm B}$.  At low baryon abundance ($\eta_{10}~\la 3$), 
a faster expansion leaves less time for \7li destruction and the relic 
abundance of mass-7 increases.  In contrast, at high baryon abundance 
($\eta_{10}~\ga 3$), $S > 1$ leaves less time for $^{7}$Be production 
and the relic abundance of mass-7 decreases.  As for D and \3he, the 
quantitative change in the \7li abundance is small for small deviations 
from SBBN.

For similar reasons, \Yp is sensitive to an asymmetry in the electron
neutrinos which, through the charged current weak interactions, help
to regulate the n/p ratio.  For $\xi_{e} > 0$, there are more neutrinos
than antineutrinos, so that reactions such as $n+\nu_{e} \longrightarrow 
p+e^{-}$, drive down the n/p ratio.  For small asymmetry at fixed 
$\eta_{\rm B}$, KS find $\Delta$Y$_{\rm P} \approx -0.23\xi_{e}$.  
The primordial abundances of D, \3he, and \7li, while not entirely 
insensitive to neutrino degeneracy, are much less affected by a 
nonzero $\xi_{e}$ than is \4he (\eg Kang \& Steigman 1992\cite{kang}).  

Each of these nonstandard cases ($S\neq 1$, $\xi \neq 0$) will be 
considered below.  While certainly not exhaustive of the nonstandard 
models proposed in the literature, they actually have the potential 
to provide semi-quantitative, if not quantitative, understanding of 
BBN in a large class of nonstandard models.  Note that data constraining 
the primordial abundances of at least two different relic nuclei (one 
of which should be \4he) are required to break the degeneracy between 
the baryon density and the additional parameter resulting from new 
physics or cosmology.  \4he is a poor baryometer but a very good 
chronometer and/or, leptometer; D, \3he, \7li have the potential to 
be good baryometers.

\subsection{Simple -- But Accurate -- Fits To The Primordial Abundances}
\label{bbnforped}

While BBN involves only a limited number of coupled differential equations,
they are non-linear and not easily solved analytically.  As a result,
detailed comparisons of the theoretical predictions with the inferred
relic abundances of the light nuclei requires numerical calculations,
which may obscure key relations between abundances and parameters, as 
well as the underlying physics.  In particular, the connection between 
the cosmological parameter set \{$\eta_{\rm B}$, $S$, $\xi_{e}$\} and 
the abundance data set \{$y_{\rm D}$, Y$_{\rm P}$, $y_{\rm Li}$\}\footnote{\Yp 
is the \4he {\it mass fraction} while the other abundances are measured 
by {\it number} compared to hydrogen.  For numerical convenience, \yd 
$\equiv 10^{5}$(D/H) and \yli $\equiv 10^{10}$(Li/H).} may be blurred, 
especially when attempting to formulate a quantitative understanding 
of how the latter constrains the former.  However, it is clear from 
Figure~\ref{fig:schrplot} that the relic, light nuclide abundances are 
smoothly varying, monotonic functions of $\eta_{\rm B}$ over a limited 
but substantial range.  While the BBN-predicted primordial abundances 
are certainly {\it not} linearly related to the baryon density (nor to 
the other parameters $S$ and $\xi_{e}$), over the restricted ranges 
identified above, KS\cite{ks} found linear fits to the predicted abundances 
(or, to powers of them) which work very well indeed.  Introducing them 
here enables and simplifies the comparison of theory with data (below) 
and permits a quick, reasonably accurate, back of the envelope, 
identification of the successes of and challenges to BBN.

\begin{figure}
\centerline{\psfig{file=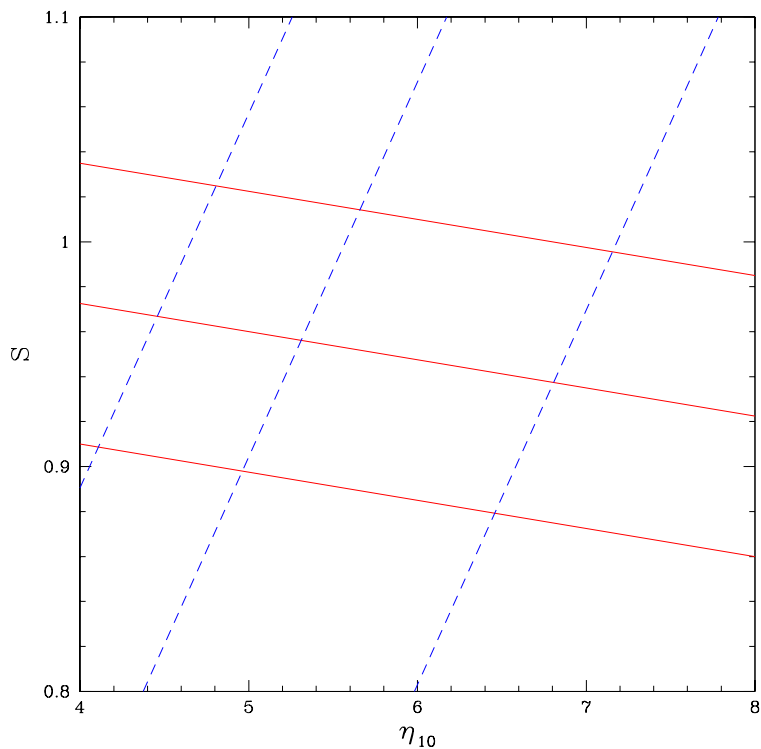,width=10.7cm}}%12cm}}
\vspace*{0pt}%\vspace*{8pt}
\caption{Isoabundance curves for Deuterium (dashed lines) and Helium-4 
(solid lines) in the expansion rate factor ($S$) -- baryon abundance 
($\eta_{10}$) plane.  The \4he curves, from bottom to top, are for 
\Yp = 0.23, 0.24, 0.25.  The D curves, from left to right, are for
\yd = 4.0, 3.0, 2.0.}
\label{fig:svseta}
\end{figure}

\begin{figure}
\centerline{\psfig{file=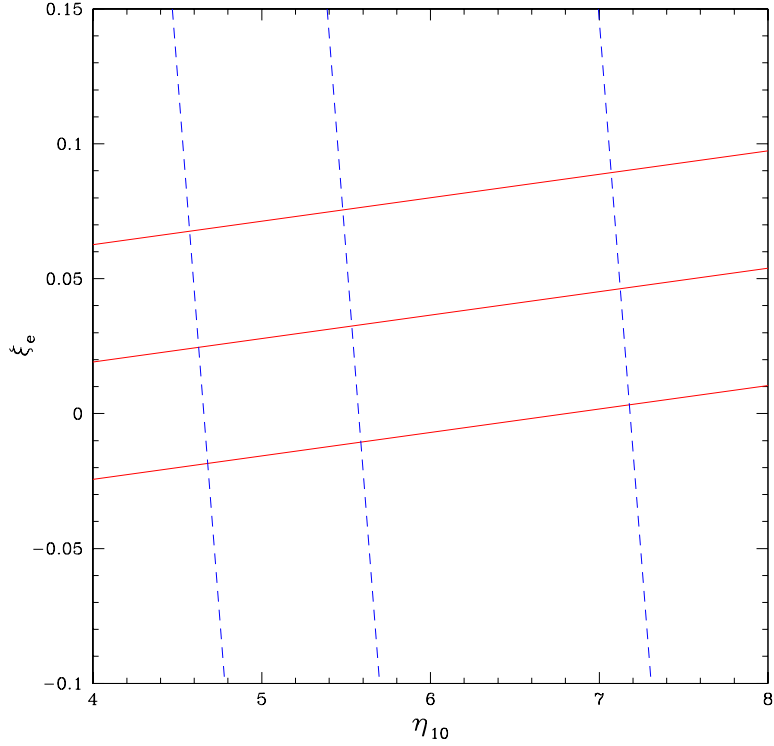,width=10.7cm}}%12cm}}
\vspace*{0pt}%\vspace*{8pt}
\caption{As in Figure~\ref{fig:svseta}, in the neutrino asymmetry ($\xi_{e}$)
-- baryon abundance ($\eta_{10}$) plane.  The \4he curves, from bottom to 
top, are for \Yp = 0.25, 0.24, 0.23.  The D curves, from left to right, 
are for \yd = 4.0, 3.0, 2.0.}
\label{fig:xivseta}
\end{figure}

For the adopted range of $\eta_{\rm B}$, \yd $= y_{\rm D}(\eta_{\rm B}$) 
is well fit by a power law,
\be 
y_{\rm D}^{FIT} \equiv 46.5\eta_{10}^{-1.6}. 
\label{ydvseta}
\ee 
While the true \yd -- $\eta_{\rm B}$ relation is not precisely a power 
law, this fit (for $4 ~\la \eta_{10} ~\la 8$) is accurate (compared to 
a numerical calculation) to better than 1\%, three times smaller than 
the $\sim 3$\% BBN uncertainty estimated by Burles, Nollett, Turner 
2001 (BNT)\cite{bnt}; this fit and the numerical calculation agree with 
the BNT result to 2\% or better over the adopted range in $\eta_{\rm B}$.  
Note that since different BBN codes are largely independent and often use 
somewhat different nuclear reaction data sets, the {\it differences} 
among their predicted abundances may provide estimates of the overall 
uncertainties.  It is convenient to introduce a ``deuterium baryon 
density parameter" $\eta_{\rm D}$, the value of $\eta_{10}$ corresponding 
to an observationally determined primordial D abundance.  
\be 
\eta_{\rm D} \equiv ({46.5(1\pm0.03) \over y_{\rm D}})^{1/1.6}. 
\ee 
Generalizing this to include the two other parameters, KS
find
\be 
\eta_{\rm D} = \eta_{10} - 6(S-1) + {5\xi_{e} \over 4}. 
\label{etad}
\ee 
This fit works quite well for $2~\la y_{\rm D}~\la 4$, corresponding 
to $5~\la \eta_{\rm D} ~\la 7$.  In Figure~\ref{fig:svseta} the deuterium
isoabundance curves are shown in the $S-\eta_{10}$ plane, while Figure~\ref
{fig:xivseta} shows the same isoabundance contours in the $\xi_{e}-\eta_{10}$
plane.  It is clear from Figures~\ref{fig:svseta} and~\ref{fig:xivseta} 
that D is a sensitive {\it baryometer} since, for these ranges of $S$ and 
$\xi_{e}$, $\eta_{\rm D} \approx \eta_{10}$.
  
Next, consider \4he.  While over a much larger range in $\eta_{10}$,
\Yp varies nearly {\it logarthmically} with the baryon density parameter, 
a {\it linear} fit to the \Yp versus $\eta_{10}$ relation is actually 
remarkably accurate over the restricted range considered here.
\be 
Y_{\rm P}^{FIT} \equiv 0.2384 + 0.0016\eta_{10} = 0.2384 + \eta_{10}/625. 
\ee 
Over the same range in $\eta_{10}$ this fit agrees with the numerical
calculation and with the BNT\cite{bnt} predictions for \Yp to within 
0.0002 ($\la 0.1$\%), or better.  Any differences between this fit 
and independent, numerical calculations are smaller (much smaller) 
than current estimates of the errors in the observationally inferred 
primordial value of Y$_{\rm P}$.  The following linear fits, including 
the total error estimate, to the \Yp -- $S$ and \Yp -- \xie relations 
from KS work very well over the adopted parameter ranges (see 
Figures~\ref{fig:svseta} \& \ref{fig:xivseta}). 
\be  
Y_{\rm P}^{FIT} \equiv 0.2384\pm 0.0006 + 0.0016\eta_{10} +  
0.16(S-1) - 0.23\xi_{e}. 
\label{yfit} 
\ee
As an aside, the dependence of the \4he mass fraction on the neutron 
lifetime ($\tau_{n}$) can be included in eq.~\ref{yfit} by adding a term 
$0.0002(\tau_{n} - 887.5)$, where $\tau_{n}$ is in seconds.  A very recent, 
new measurement of $\tau_{n}$ by Serebrov {\it et al.}\cite{taun} suggests 
that the currently accepted value ($\tau_{n} = 887.5$~s) should be 
reduced by 7.2~s.  If confirmed, this would lead to a slightly smaller 
BBN-predicted \4he abundance: $\Delta$Y$_{\rm P} = -0.0014$.  The 
corresponding shift in the \4he inferred baryon density parameter 
is negligible compared to its range of uncertainty ($\Delta\eta_{\rm 
B}/\eta_{\rm B} = -0.14$), as is that for the shift in the upper bound 
to N$_{\nu}$ ($\Delta$N$_{\nu}^{max} = +0.11$).  These corrections are 
ignored here.

In analogy with the deuterium baryon density parameter introduced
above, it is convenient to introduce $\eta_{\rm He}$, defined by
\be 
\eta_{\rm He} \equiv 625({\rm Y}_{\rm P} - 0.2384\pm0.0006), 
\ee 
so that 
\be  
\eta_{\rm He} = \eta_{10} + 100(S-1) - {575\xi_{e} \over 4}. 
\label{eq:etahe} 
\ee 
For SBBN ($S = 1$ \& \xie = 0), $\eta_{\rm He}$ is the value of $\eta_{10}$ 
corresponding to the adopted value of Y$_{\rm P}$.  Once \Yp is chosen, 
the resulting value of $\eta_{\rm He}$ provides a {\it linear} constraint  
on the combination of $\eta_{10}$, $S$, and \xie in eq.~\ref{eq:etahe}.   
This fit works well\cite{ks} for $0.23~\la $Y$_{\rm P}~\la 0.25$, 
corresponding to $-5~\la \eta_{\rm He}~\la 7$.  As Figures~\ref{fig:svseta} 
\& \ref{fig:xivseta} reveal, \4he is an excellent chronometer and/or 
leptometer, since the \Yp isoabundance curves are nearly horizontal 
(and very nearly orthogonal to the deuterium isoabundance curves).
 
As with D, the \7li abundance\footnote{It is common in the astronomical  
literature to present the lithium abundance logarithmically: [Li]~$\equiv  
12 + $log(Li/H) = 2 + log(y$_{\rm Li}$).} is well described by a power  
law in $\eta_{10}$ over the range in baryon abundance explored here: 
\yli $\equiv 10^{10}$(Li/H) $\propto \eta_{10}^{2}$.  The following KS 
fit agrees with the BBN predictions to better than 3\% over the adopted 
range in $\eta_{10}$, 
\be 
y_{\rm Li}^{FIT} \equiv {\eta_{10}^{2} \over 8.5}. 
\ee 
While this fit predicts slightly smaller lithium abundances compared  
to those of BNT\cite{bnt}, the differences are at the 5-8\% level, small 
compared to the BNT uncertainty estimates as well as those of Hata 
{\it et al.} (1995)\cite{hata} ($\sim 10 - 20$\%). 
 
In analogy with $\eta_{\rm D}$ and $\eta_{\rm He}$ defined above, the 
lithium baryon abundance parameter $\eta_{\rm Li}$ (allowing for a 10\% 
overall uncertainty) is defined by 
\be 
\eta_{\rm Li} \equiv (8.5(1\pm0.1)y_{\rm Li})^{1/2}. 
\ee 
The simple, linear relation for $\eta_{\rm Li}$ as a function  
of $\eta_{10}$, $S$, $\xi_{e}$, which KS find fits reasonably well 
over the adopted parameter ranges is, 
\be 
\eta_{\rm Li} = \eta_{10} - 3(S-1) - {7\xi_{e} \over 4}. 
\label{eq:etali} 
\ee 
This fit works well for $3~\la y_{\rm Li}~\la 5$, corresponding 
to $5~\la \eta_{\rm Li}~\la 7$, but it breaks down for \yli $\la 2$ 
($\eta_{\rm Li} ~\la 4$); see Fig.~\ref{fig:schrplot}.  As is the 
case for deuterium, lithium can be an excellent baryometer since, 
for the restricted ranges of $S$ and $\xi_{e}$ under consideration 
here, $\eta_{\rm Li} \approx \eta_{10}$. 

Finally, it may be of interest to note that for \3he the power law 
$y_{3} - \eta_{\rm B}$ relation, where $y_{3} \equiv 10^{5}($\3he/H), 
which is reasonably accurate for $4~\la \eta_{10}~\la 8$ is
\be
y_{3} = 3.1(1\pm0.03)\eta_{10}^{-0.6}.
\ee
The difficulty of using current observational data, limited to 
chemically evolved regions of the Galaxy, to infer the primordial 
abundance of \3he, along with the relatively weak dependence of 
$y_{3}$ on $\eta_{10}$, limits the utility of this nuclide as a 
baryometer\cite{3he}.  \3he can, however, be used as a test of BBN 
consistency.

\subsection{SBBN-Predicted Primordial Abundances}
\label{sbbn}

Before discussing the current status of the observationally determined
abundances (and their uncertainties) of the light nuclides, it is
interesting to {\it assume} SBBN and, for the one free parameter, 
$\eta_{\rm B}$, use the value inferred from non-BBN data such as the 
CBR (WMAP) and Large Scale Structure (LSS)\cite{sperg} to predict the 
relic abundances.

From WMAP alone, Spergel {\it et al.} 2003\cite{sperg} derive $\eta_{10} 
= 6.3 \pm 0.3$.  Using the fits from \S\ref{bbnforped}, with $S=1$ and 
\xie = 1, the SBBN-predicted relic abundances are: \yd $= 2.45 \pm 0.20$; 
$y_{3} = 1.03 \pm 0.04$; \Yp $= 0.2485 \pm 0.0008$; \yli $= 4.67 \pm 0.64$ 
([Li]$_{\rm P} = 2.67 \pm 0.06$).

When Spergel {\it et al.} 2003\cite{sperg} combine the WMAP CBR data with 
those from Large Scale Structure, they derive a consistent, but slightly
smaller (slightly more precise) baryon abundance parameter $\eta_{10} 
= 6.14 \pm 0.25$.  For this choice the SBBN-predicted relic abundances 
are: \yd $= 2.56 \pm 0.18$; $y_{3} = 1.04 \pm 0.04$; \Yp $= 0.2482 \pm 
0.0007$; \yli $= 4.44 \pm 0.57$ ([Li]$_{\rm P} = 2.65^{+0.05}_{-0.06}$).

\section{Observationally Inferred Primordial Abundances}
\label{abund}
  
\subsection{The Primordial Deuterium Abundance}
\label{d}
\begin{figure}
\centerline{\psfig{file=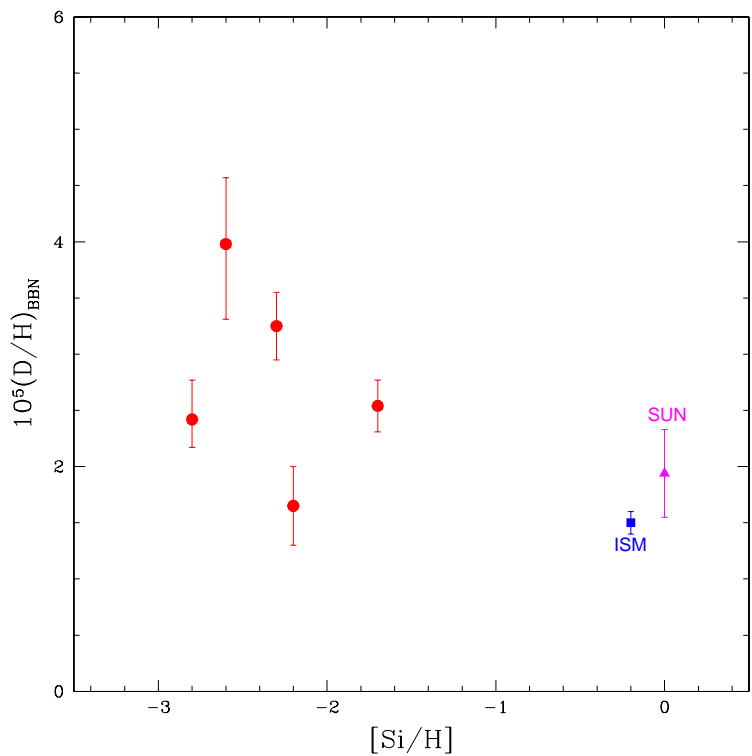,width=10.7cm}}%12cm}}
\vspace*{0pt}%\vspace*{8pt}
\caption{The observationally inferred primordial deuterium abundances
(the ratio of D to H by number) versus a logarithmic measure of the 
metallicity, relative to solar ([Si/H]), for five high redshift, low 
metallicity QSOALS (filled circles) through 2003.  The error bars are 
the quoted $1\sigma$ uncertainties.  Also shown for comparison are the 
D/H ratios inferred from observations of the local interstellar medium 
(ISM; filled square) and that for the pre-solar nebula (Sun; filled 
triangle).}
\label{fig:dvssi03}
\end{figure}

Deuterium is the baryometer of choice since its post-BBN evolution is 
simple (and monotonic!) and its BBN-predicted relic abundance depends 
sensitively on the baryon abundance (\yd $\propto \eta_{\rm B}^{-1.6}$).  
As the most weakly bound of the light nuclides, any deuterium cycled through 
stars is burned to \3he and beyond during the pre-main sequence, convective 
(fully mixed) evolutionary stage\cite{els}.  Thus, deuterium observed anywhere, 
anytime, should provide a {\it lower} bound to the primordial D abundance.  
For ``young'' systems at high redshift and/or with very low metallicity, 
which have experienced very limited stellar evolution, the observed D 
abundance should be close to the primordial value.  Thus, although there 
are observations of deuterium in the solar system and the interstellar
medium (ISM) of the Galaxy which provide interesting {\it lower} bounds 
to the primordial abundance, it is the observations of relic D in a 
few (too few!), high redshift, low metallicity, QSO absorption line 
systems (QSOALS) which are of most value in enabling estimates of 
its primordial abundance.  

\begin{figure}
\centerline{\psfig{file=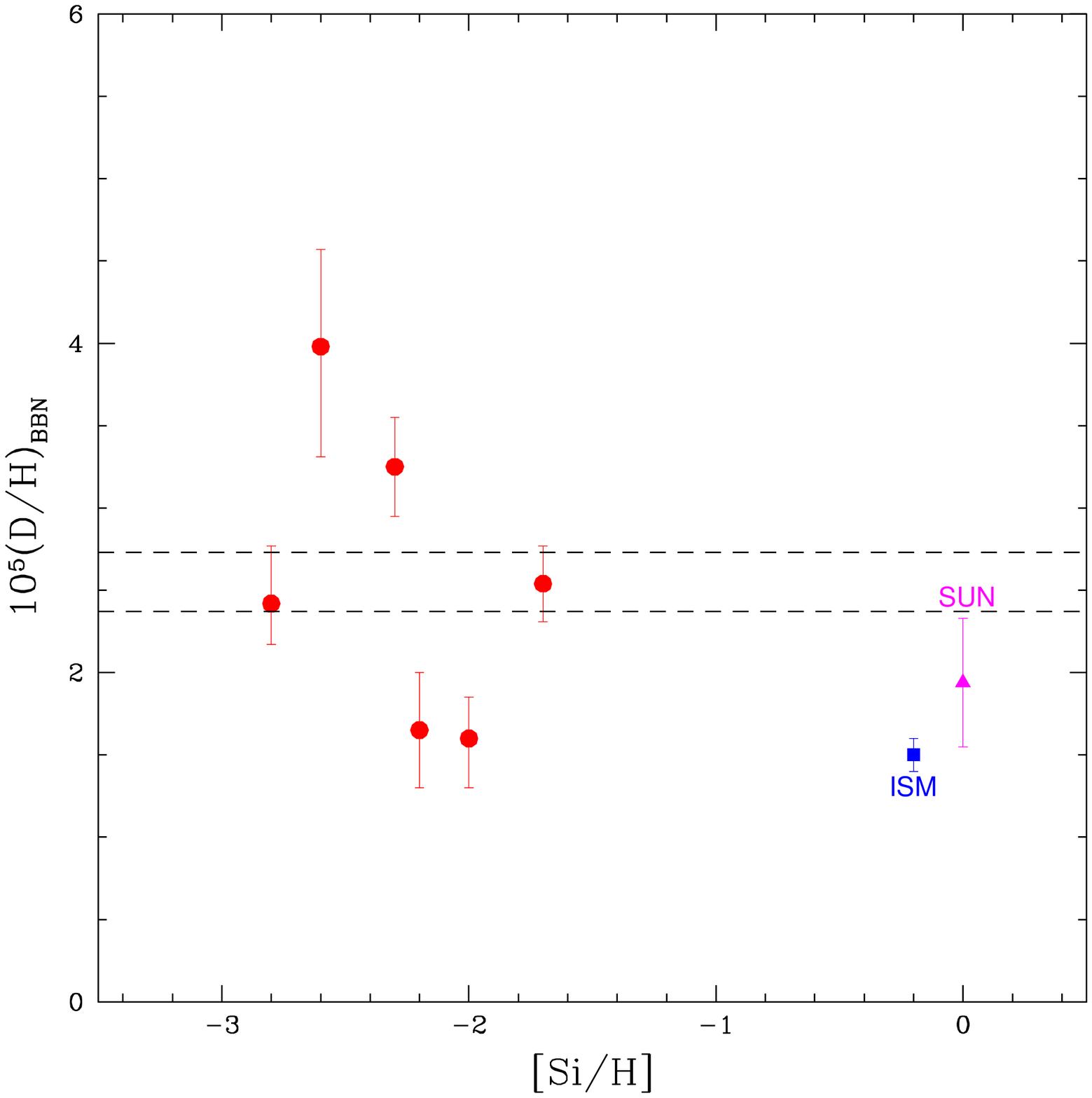,width=10.7cm}}%12cm}}
\vspace*{0pt}%\vspace*{8pt}
\caption{Figure~\ref{fig:dvssi03} updated to 2004 to include the one 
new deuterium abundance determination$^{18}$ for a high redshift, 
low metallicity QSOALS.  The dashed lines show the SBBN-predicted 
$1\sigma$ band for the WMAP baryon abundance.}
\label{fig:dvssi04}
\end{figure}

In contrast to the great asset of the simple post-BBN evolution, the 
identical absorption spectra of \di and \hi (modulo the velocity/wavelength 
shift resulting from the heavier reduced mass of the deuterium atom) 
is a severe liability, limiting drastically the number of useful targets 
in the vast Lyman-alpha forest of QSO absorption spectra (see Kirkman 
{\it et al.}\cite{kirk} for further discussion).  As a result, it is essential 
to choose target QSOALS whose velocity structure is ``simple'' since 
a low column density \hi absorber, shifted by $\sim 81$~km/s with respect 
to the main \hi absorber (an ``interloper'') could masquerade as \di 
absorption\cite{interloper}.  If this degeneracy is not recognized, a D/H 
ratio which is too high could be inferred.  Since there are many more 
low-column 
density absorbers than those with high \hi column densities, absorption 
systems with somewhat lower \hi column density (\eg Lyman-limit systems: 
LLS) may be more susceptible to this contamination than the higher \hi 
column density absorbers (\eg damped Ly$\alpha$ absorbers: DLA).  While 
the DLA do have many advantages over the LLS, a precise determination 
of the \hi column density utilizing the damping wings of the \hi absorption 
requires an accurate placement of the continuum, which could be compromised 
by \hi interlopers.  This might lead to errors in the \hi column 
density.  These complications are real and the path to primordial 
D using  QSOALS  has not been straightforward, with some abundance claims 
having had to be withdrawn or revised.  Through 2003 there were only five 
``simple" QSOALS with deuterium detections leading to reasonably robust
abundance determinations\cite{kirk} (and references therein); these are 
shown in Figure~\ref{fig:dvssi03} along with the corresponding solar system 
and ISM D abundances.  It is clear from Figure~\ref{fig:dvssi03}, that there 
is significant dispersion among the derived D abundances at low metallicity 
which, so far, masks the anticipated primordial deuterium plateau, suggesting 
that systematic (or random) errors, whose magnitudes are hard to estimate, 
may have contaminated the determinations of the \di and/or \hi column 
densities. 

It might be hoped that as more data are acquired, the excessive dispersion
among the deuterium abundances seen in Fig.~\ref{fig:dvssi03} would decrease 
leading to a better defined deuterium plateau.  Be careful what you wish 
for!  In 2004, Crighton {\it et al.}\cite{webb} identified deuterium absorption 
in another high redshift, low metallicity QSOALS and derived its abundance. 
The updated version of Fig.~\ref{fig:dvssi03} is shown in Fig.~\ref{fig:dvssi04}. 
Now, the dispersion is {\bf larger}!  The low value of the Crighton {\it et 
al.}\cite{webb} D abundance, similar to that in the more highly evolved ISM 
and the pre-solar nebula, is puzzling.  D.~Tytler (private communication) 
and colleagues have data (unpublished) for the same QSOALS and they find 
acceptable fits with lower \hi column densities and no visible 
D\thinspace{$\scriptstyle{\rm I}$}; perhaps this system is contaminated by an 
interloper\cite{interloper}.  The two low D/H ratios, if not affected by 
random or systematic errors, may be artifacts of a small statistical sample 
or, they may have resulted from "young" regions in which some of the relic 
deuterium has been destroyed or depleted onto dust\cite{draine}.  These
suggestions pose challenges since any cycling of gas through stars should 
have led to an {\it increase} in the heavy element abundances (these QSOALS 
are metal poor) and, at low metallicity the amount of dust is expected to 
be small. 

For the Spergel {\it et al.}\cite{sperg} baryon abundance of $\eta_{10} = 
6.14 \pm 0.25$, the SBBN-predicted deuterium abundance is \yd = $2.6 \pm 
0.2$.  As may be seen in Fig.~\ref{fig:dvssi04}, the current data exhibit 
the Goldilocks effect: two D/H ratios are ``too small", two are ``too 
large", and two are ``just right".  Nonetheless, it is clear that the 
sparse data currently available are in very good agreement with the SBBN 
prediction.  If the weighted mean of D/H for the six QSOALS is adopted, 
but the dispersion in the mean is used in place of the error in the mean, 
\yd = $2.4 \pm 0.4$, corresponding to an SBBN-predicted baryon abundance 
of $\eta_{10} = 6.4 \pm 0.6$, in excellent agreement with the WMAP value.  
Were it not for the excessive dispersion among the extant deuterium 
abundance determinations, the precision of the baryon abundance determined 
from SBBN would be considerably enhanced.  For example, if the formal 
error in the mean is used, $\eta_{10} = 6.4 \pm 0.2$ is the deuterium-based, 
SBBN prediction, very close to that for the baryon abundance derived 
from the CBR alone\cite{sperg} ($\eta_{10}^{\rm WMAP} = 6.3 \pm 0.3$).
More data is crucial to deuterium fulfilling its potential.

\begin{figure}
\centerline{\psfig{file=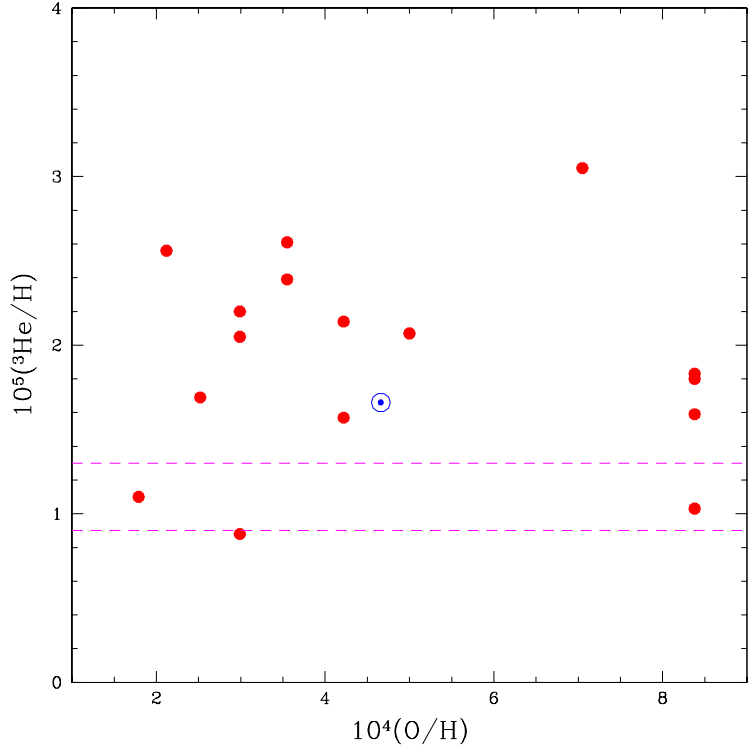,width=10.7cm}}%12cm}}
\vspace*{0pt}%\vspace*{8pt}
\caption{The \3he abundance determinations (by number relative to H) in 
the ISM of the Galaxy (from BRB$^{24}$) as a function of the corresponding 
oxygen abundances.  The solar symbol indicates the \3he abundance for 
the pre-solar nebula.  The dashed lines show the $1\sigma$ band adopted 
by BRB.}
\label{fig:3hevso}
\end{figure}

\subsection{The Primordial Helium-3 Abundance}
  
\begin{figure}
\centerline{\psfig{file=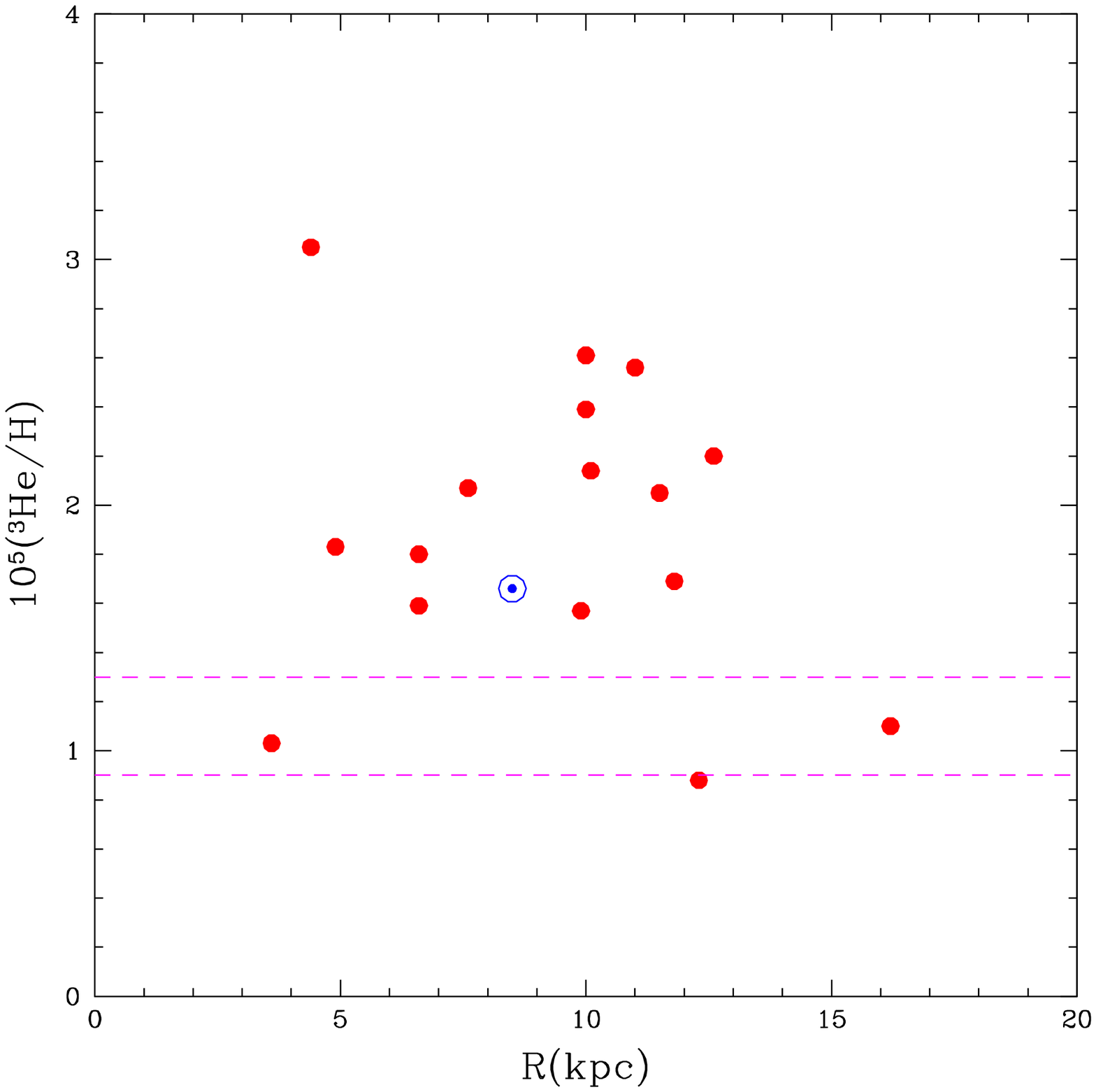,width=10.7cm}}%12cm}}
\vspace*{0pt}%\vspace*{8pt}
\caption{As in Figure~\ref{fig:3hevso} but, for the \3he abundances 
as a function of distance from the center of the Galaxy.}
\label{fig:3hevsr}
\end{figure}

The post-BBN evolution of \3he, involving competition among stellar 
production, destruction, and survival, is considerably more complex 
and model dependent than that of D.  Interstellar \3he incorporated 
into stars is burned to \4he (and beyond) in the hotter interiors, 
but it is preserved in the cooler, outer layers.  Furthermore, while 
hydrogen burning in cooler, low-mass stars is a net producer of 
\3he\cite{3hestars}, it is unclear how much of this newly synthesized 
\3he is returned to the interstellar medium and how much of it is 
consumed in post-main sequence evolution (\eg Sackmann \& Boothroyd\cite{sb}).  
For years it had been anticipated that net stellar production would 
prevail in this competition, so that the \3he abundance would increase 
with time (and with metallicity)\cite{3heevol}.  

Observations of \3he, are restricted to the solar system and the Galaxy.  
Since for the latter there is a clear gradient of metallicity with location, 
a gradient in \3he abundance was also expected.  However, as is clear 
from Figures~\ref{fig:3hevso} \& \ref{fig:3hevsr}, the data\cite{gg,brb} 
reveal no statistically significant correlation between the \3he 
abundance and metallicity or location in the Galaxy, suggesting a 
very delicate balance between net production and net destruction of 
\3he.  For a recent review of the current status of \3he evolution, 
see Romano {\it et al.}\cite{romano}.  

While the absence of a gradient suggests the {\it mean} (``plateau") 
\3he abundance in the Galaxy ($y_{3} \approx 1.9 \pm 0.6$) might 
provide a good estimate of the primordial abundance, Bania, Rood \& 
Balser (BRB)\cite{brb} prefer to adopt as an upper limit to the primordial 
abundance, the \3he abundance measured in the most distant (from the 
Galactic center), most metal poor, Galactic \hii region, $y_{3}~\la 
1.1\pm0.2$; see Figs. \ref{fig:3hevso} \& \ref{fig:3hevsr}.  This choice 
is in excellent agreement with the SBBN/WMAP predicted abundance of 
$y_{3} = 1.04 \pm 0.04$ (see \S\ref{bbnforped}).  While both D and \3he 
are consistent with the SBBN predictions, \3he is a less sensitive 
baryometer than is D since (D/H)$_{\rm BBN} \propto \eta_{\rm B}^{-1.6}$, 
while $(^{3}$He/H)$_{\rm BBN} \propto \eta_{\rm B}^{-0.6}$.  For example, 
if $y_{3} = 1.1 \pm 0.2$ is adopted for the \3he primordial abundance, 
$\eta_{10}(^{3}$He$) = 6.0 \pm 1.7$.  While the central value of the 
\3he-inferred baryon density parameter is in nearly perfect agreement 
with the WMAP value\cite{sperg}, the allowed range of $\eta_{\rm B}$ is 
far too large to be very useful.  Still, \3he can provide a valuable 
BBN consistency check.

\subsection{The Primordial Helium-4 Abundance}
\label{4he}

\begin{figure}
\centerline{\psfig{file=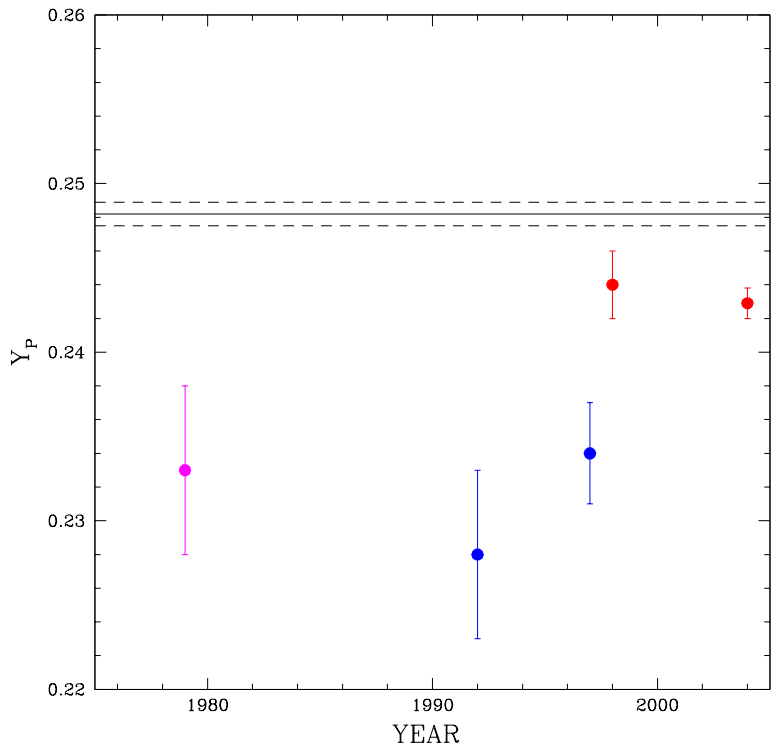,width=10.7cm}}%12cm}}
\vspace*{0pt}%\vspace*{8pt}
\caption{The observationally inferred primordial \4he mass fractions
from 1978 until 2004.  The error bars are the quoted $1\sigma$ 
uncertainties.  Also shown is the SBBN-predicted relic abundance 
(solid line) for the WMAP baryon abundance, along with the $1\sigma$ 
uncertainty (dashed lines) of the SBBN prediction.}
\label{fig:hevstime}
\end{figure}

The post-BBN evolution of \4he is quite simple.  As gas cycles 
through generations of stars, hydrogen is burned to helium-4 
(and beyond), increasing the \4he abundance above its primordial 
value.  The \4he mass fraction in the Universe at the present 
epoch, Y$_{0}$, has received a significant contribution from 
post-BBN, stellar nucleosynthesis, so that Y$_{0} >$~Y$_{\rm P}$.  
However, since the ``metals" such as oxygen are produced by 
short-lived, massive stars and \4he is synthesized (to a greater 
or lesser extent) by all stars, at very low metallicity the 
increase in Y should lag that in \eg O/H so that as O/H 
$\rightarrow 0$, Y $\rightarrow$ Y$_{\rm P}$.  As is the case 
for deuterium and lithium, a \4he ``plateau'' is expected at 
sufficiently low metallicity.  Therefore, although \4he is 
observed in the Sun and in Galactic \hii regions, the key 
data for inferring its primordial abundance are provided by 
observations of helium and hydrogen emission (recombination) 
lines from low-metallicity, extragalactic \hii regions.  The 
present inventory of such regions studied for their helium 
content exceeds 80 (see Izotov \& Thuan (IT)\cite{it}).  Since 
with such a large data set even modest observational errors for 
the individual \hii regions can lead to an inferred primordial 
abundance whose formal {\it statistical} uncertainty may be quite 
small, special care must be taken to include hitherto ignored or 
unaccounted for {\it systematic} corrections and/or errors.  It is the 
general consensus that the present uncertainty in \Yp is dominated 
by the latter, rather than by the former errors.  Indeed, many of
the potential pitfalls were identified by Davidson \& Kinman\cite{dk} 
in a prescient, 1985 paper.  In the abstract they say, ``The most 
often quoted estimates of the primordial helium abundance are 
optimistic in the sense that quoted uncertainties usually do 
not include some potentially serious systematic errors."  

To provide a context for the discussion of the most recent data 
and analyses, Figure~\ref{fig:hevstime} offers a compilation of 
the history of \Yp determinations\cite{it,hevstime} derived using 
data from low metallicity, extragalactic \hii regions.  Notice 
that all of these estimates, taken at face value, fall below 
the SBBN/WMAP predicted primordial abundance by at least 
2$\sigma$, reemphasizing the importance of accounting for 
systematic uncertainties.  With this in mind, we turn to recent 
reanalyses\cite{os,peim} of the IT data\cite{it}, supplemented by key 
observations of a local, higher metallicity \hii region\cite{peim}.

\begin{figure}
\centerline{\psfig{file=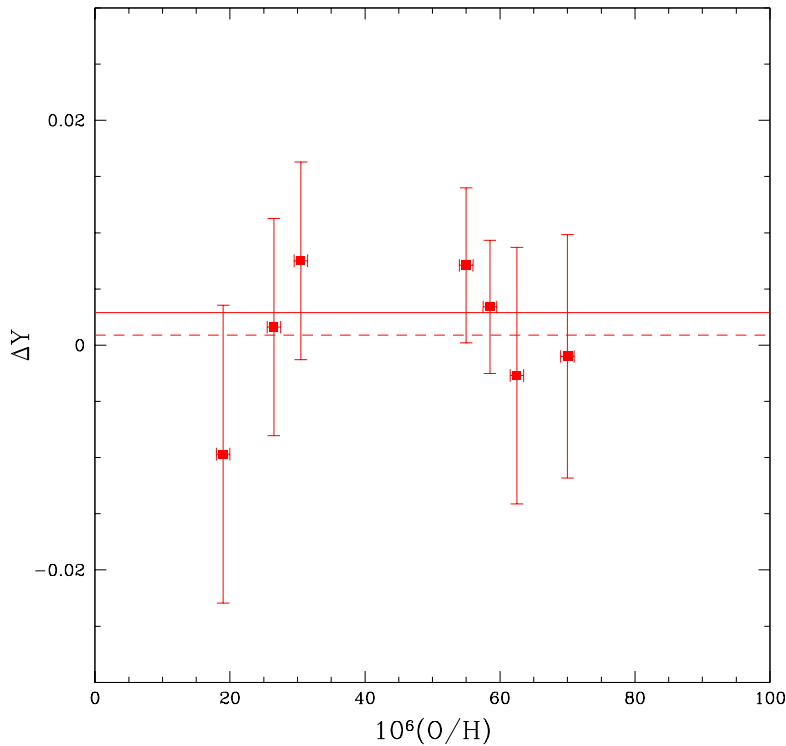,width=10.7cm}}%12cm}
\vspace*{0pt}%\vspace*{8pt}
%\vspace*{8pt}
\caption{The {\bf{\it differences}} between the OS and IT \4he abundances, 
$\Delta$Y$ \equiv $Y$^{\rm OS} - $Y$^{\rm IT}$ for the OS-selected IT \hii 
regions versus the corresponding oxygen abundances.  The solid line is 
the weighted mean of the helium mass fraction differences, while the 
dashed line shows the unweighted {\it average} of the differences.}
\label{fig:delyvso}
\end{figure}

Prior and subsequent to the Davidson \& Kinman paper\cite{dk} astronomers
have generally been aware of the important sources of potential
systematic errors associated with using recombination line data
to infer the helium abundance.  However, attempts to account for
them have often been unsystematic or, entirely absent.  The current
conventional wisdom that the accuracy of the data demands their
inclusion has led to some attempts to account for a few of them 
or, for combinations of a few of them\cite{it,os,peim,svg,icf}.  
The Olive \& Skillman (OS)\cite{os} analysis of the IT data is the 
most systematic to date.  Following criteria outlined in their 
2001 paper\cite{os}, OS found they were able to apply their analysis 
to only 7 of the 82 IT \hii regions.  This tiny data set, combined with 
its limited range in metallicity (oxygen abundance), severely limits 
the statistical significance of any conclusions OS can extract from 
it.  In Figure~\ref{fig:delyvso} are shown the {\it differences} between 
the OS-inferred and the IT-inferred helium abundances.  For these 
seven \hii regions there is no evidence that $\Delta {\rm Y} \equiv 
{\rm Y}^{\rm OS}-{\rm Y}^{\rm IT}$ is correlated with metallicity.  
The weighted mean offset along with the error in the mean are 
$\Delta {\rm Y} = 0.0029 \pm 0.0032$ (the {\it average} offset and 
the {\it average} error are $\Delta {\rm Y} = 0.0009 \pm 0.0095$), 
consistent with zero at 1$\sigma$.

\begin{figure}
\centerline{\psfig{file=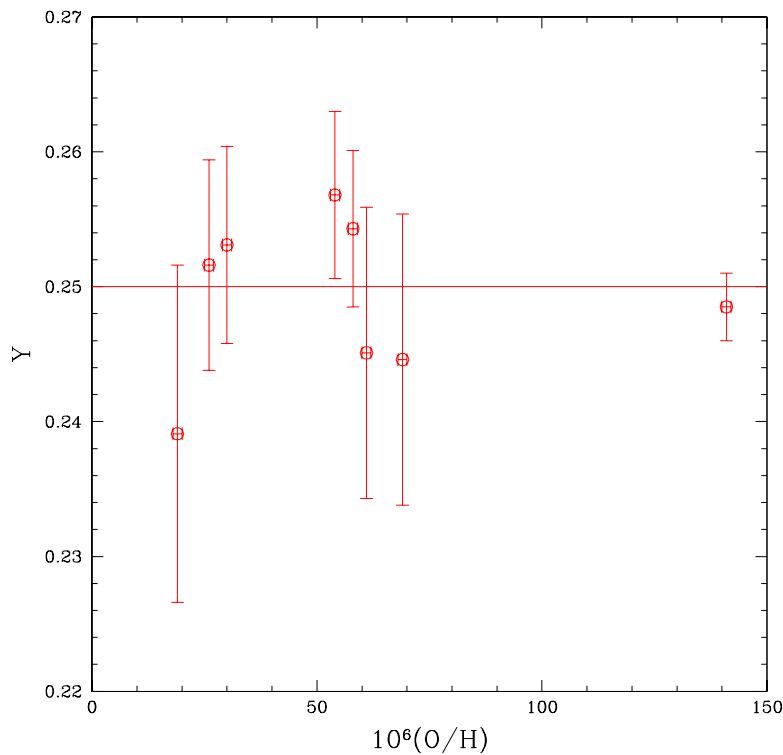,width=10.7cm}}%12cm}}
\vspace*{0pt}%\vspace*{8pt}
\caption{The OS-revised \4he versus oxygen abundances for the seven 
IT \hii regions and the SMC \hii region from PPR.  The solid line 
is the weighted mean of the helium abundances for all eight of 
the \hii regions reanalyzed by OS.}
\label{fig:oshevso}
\end{figure}

If the weighted mean offset is applied to the IT-derived primordial 
abundance of Y$^{\rm IT}_{\rm P} = 0.2443 \pm 0.0015$, the ``corrected" 
primordial value becomes
\be
{\rm Y}^{\rm OS}_{\rm P} \equiv {\rm Y}^{\rm IT}_{\rm P} + \Delta {\rm Y}
= 0.2472 \pm 0.0035,
\label{osyp}
\ee
leading to a 2$\sigma$ {\it upper} bound on the primordial abundance of 
Y$^{\rm OS}_{\rm P} \leq 0.254$.  In contrast, OS prefer to fit these 
seven data points to a {\it linear} Y versus O/H relation and, from it, 
derive the primordial abundance.  Their IT-revised abundances, along 
with, for comparison, that from their reanalysis of the Peimbert 
{\it et al.}\cite{peim} data for an \hii region in the SMC (to be discussed 
next), are shown in Figure~\ref{fig:oshevso}.  It is not surprising that 
for only seven data points, each with larger errors than those adopted 
by IT, spanning such a narrow range in metallicity, their linear fit, 
Y$^{\rm OS}_{7} = 0.2495 \pm 0.0092 + (54 \pm 187)$(O/H), is {\bf not} 
statistically significant.  Indeed, it is {\bf not} preferred over the 
simple weighted mean of the seven helium abundances ($0.252 \pm 0.003$), 
since the $\chi^2$ per degree of freedom is actually {\it higher} for the 
linear fit.  In fact, there is no statistically significant correlation 
between Y and O/H for the IT-derived abundances for these seven \hii 
regions either.  As valuable as is their reanalysis of the IT data, 
the OS conclusion that \Yp $= 0.249 \pm 0.009$ is not supported by the 
sparse data set they used\footnote{Note that OS used the corresponding 
1$\sigma$ upper bound of 0.258 for the upper bound to their ``favored" 
primordial abundance range.}.  Unless and until an analysis is performed 
of a much larger data set, with a longer metallicity baseline, the 
estimate in eq.~\ref{osyp}, and its corresponding 2$\sigma$ upper bound, 
may provide a good starting point at present for an approach to the 
primordial abundance of \4he.

Another correction, not directly constrained by the analysis of OS, is 
related to the inhomogeneous nature of \hii regions.  Unlike classical, 
textbook, homogeneous, Str$\ddot{\rm o}$mgren spheres, real \hii regions 
are filamentary and inhomogeneous, with variations in electron density 
and temperature likely produced by shocks and winds from pockets of hot, 
young stars.  Esteban and Peimbert\cite{est} have noted that temperature 
fluctuations can have a direct effect on the helium abundance derived 
from recombination lines.  This effect was investigated theoretically 
using models of \hii regions\cite{svg}, but more directly by Peimbert {\it 
et al.}\cite{peim} using data from a nearby, spatially resolved, \hii region 
in the Small Magellanic Cloud (SMC), along with their reanalyses of four 
\hii regions selected from IT.  While the SMC \hii region formed out of 
chemically evolved gas and, therefore, cannot be used by itself to derive
primordial abundances, the spatial resolution it offers permits a direct 
investigation of many potential systematic effects.  In particular, since 
recombination lines are used, the observations are blind to any neutral 
helium or hydrogen.  Estimates of the ``ionization correction factor" 
({\it icf}), while model dependent, are large\cite{icf}.  For example, using 
models of \hii regions ionized by distributions of stars of different 
masses and ages and comparing to the IT (1998) data, Gruenwald {\it et 
al.}\cite{icf} concluded that IT {\it overestimated} the primordial \4he 
abundance by $\Delta$Y$^{\rm GSV}(icf) \approx 0.006 \pm 0.002$; Sauer 
\& Jedamzik\cite{icf} find a similar, even larger, correction.  If this 
correction is applied to the OS-revised, IT primordial abundance in 
eq.~\ref{osyp}, the new, {\it icf}-corrected value is
\be
{\rm Y}^{\rm GSV}_{\rm P}(icf) \equiv {\rm Y}^{\rm OS}_{\rm P} - 
\Delta {\rm Y}^{\rm GSV}(icf) = 0.241 \pm 0.004.
\label{icfyp}
\ee

In addition to the subset of 7 of the 82 IT \hii regions which meet 
their criteria, OS also reanalyzed the Peimbert {\it et al.} data\cite{peim} 
for the SMC \hii region.  This OS-revised data point is shown in Figure
\ref{fig:oshevso} at the highest oxygen abundance.  Notice that the eight 
data points plotted in Fig.~\ref{fig:oshevso} show no evidence of the 
expected {\it increase} of Y with metallicity\footnote{Absence of evidence 
is NOT evidence of absence.}; this is likely due to the small sample 
size.  The weighted mean \4he abundance for these eight \hii regions 
is Y$^{\rm OS}_{8} = 0.250 \pm 0.002$, corresponding to a 2$\sigma$ 
upper bound of \Yp $\leq 0.254$.  Coincidentally, this is the same 
2$\sigma$ upper bound as that found from the mean of the seven IT \hii 
regions (see eq.~\ref{osyp}) and, also, the 2$\sigma$ upper bound from 
the OS-reanalyzed SMC \hii region alone.  If the ionization corrections 
from Gruenwald {\it et al.}\cite{icf} are applied to each of these eight 
\hii regions, it is found that the mean $\Delta$Y$(icf)_{8} = -0.002 
\pm 0.002$, so that including this correction, while accounting for 
the increased error, leaves the 2$\sigma$ upper bound of \Yp $\leq 
0.254$ unchanged.

The lesson from the discussion above is that while recent attempts
to determine the primordial abundance of \4he may have achieved high 
{\it precision}, their {\it accuracy} remains in question.  The latter
is limited by our understanding of and our ability to account for 
systematic errors and biases, not by the statistical uncertainties.  
The good news is that carefully organized, detailed studies of only 
a few ($\sim$~a dozen?) low metallicity, extragalactic \hii regions
may go a long way towards an accurate determination of Y$_{\rm P}$.
The bad news is that many astronomers and telescope allocation
committee members are unaware that this is an interesting and 
important problem, worth their effort and telescope time.  At
present then, the best that can be done is to adopt a defensible 
value for \Yp and, especially, its uncertainty.  To this end, in 
the following the estimate in eq.~\ref{icfyp}~is chosen: \Yp $= 0.241 
\pm 0.004$.  While the central value of \Yp is low, it is within 2$\sigma$ 
($\sim 1.75\sigma$) of the SBBN/WMAP expected central value of \Yp = 
0.248 (see \S\ref{bbnforped}).   Note that the {\it extrapolation} of 
the linear fit of the \{Y,~O/H\} data from the lowest metallicity 
(O/H $\approx 2\times 10^{-5}$) to zero metallicity (Y$_{\rm P}$) 
corresponds to $\Delta$Y $\approx 0.0009$, well within the 
uncertainties of Y$_{\rm P}$.

In setting contraints on new physics, an upper bound to \Yp is 
required.  A robust upper bound suggested by the above discussion 
is \Yp $\leq 0.254$.  As an example, the SBBN/WMAP lower bound
(at $\sim 2\sigma$) to \Yp is 0.247, so that $\Delta$Y$_{\rm P} <
0.007$.  This corresponds to the robust upper bounds $S < 1.04$
and N$_{\nu} < 3.5$, eliminating (just barely) even one, new, light
scalar, and bounding the lepton asymmetry from below: $\xi_{e} 
> -0.03$.

\subsection{The Primordial Lithium-7 Abundance}

\begin{figure}
 %\centering
  %\vspace{-7.0pc}
   %\epsfysize=13.5cm
    \vspace{-9.0pc}
    \epsfysize=13.5cm
    \epsfbox{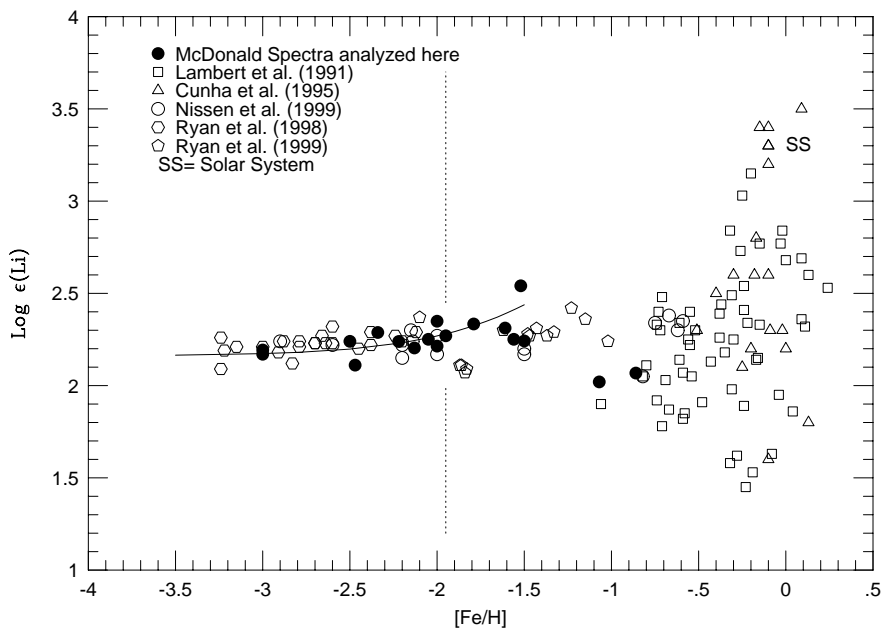}
\caption{Lithium abundances, log~$\epsilon$(Li)~$ \equiv$~[Li] $\equiv 
12 + $log(Li/H) versus metallicity (on a log scale relative to solar) 
from a compilation of stellar observations by V. V. Smith.  The solid 
line is intended to guide the eye to the ``Spite Plateau".}
\label{fig:livsfe}  
\end{figure}    

In the post-BBN universe \7li, along with $^{6}$Li, $^{9}$Be, $^{10}$B, 
and $^{11}$B, is produced in the Galaxy by cosmic ray spallation/fusion 
reactions.  Furthermore, observations of super-lithium rich red giants 
provide evidence that (at least some) stars are net producers of lithium.  
Therefore, even though lithium is easily destroyed in the hot interiors 
of stars, theoretical expectations supported by the observational data 
shown in Figure~\ref{fig:livsfe} suggest that while lithium may have been 
depleted in many stars, the overall trend is that its abundance has 
increased with time.  Therefore, in order to probe the BBN yield of \7li, 
it is necessary to restrict attention to the oldest, most metal-poor halo 
stars in the Galaxy (the ``Spite Plateau") seen at low metallicity in Fig.
\ref{fig:livsfe}.  Using a selected set of the lowest metallicity halo stars, 
Ryan {\it et al.}\cite{ryan} claim evidence for a 0.3 dex increase in the lithium 
abundance ([Li] $\equiv 12 + $log(Li/H)) for $-3.5 \leq$ [Fe/H] $\leq -1$, 
and they derive a primordial abundance of [Li]$_{\rm P} \approx 2.0-2.1$.  
This abundance is low compared to the value found by Thorburn\cite{thor}, 
who derived [Li]$_{\rm P} \approx 2.25\pm0.10$.  The stellar temperature 
scale plays a key role in using the observed equivalent widths to derive 
the \7li abundance.  Studies of halo and Galactic Globular Cluster stars 
employing the infrared flux method effective temperature scale suggest a 
higher lithium plateau abundance\cite{bonif}: [Li]$_{\rm P} = 2.24\pm0.01$, 
similar to Thorburn's\cite{thor} value.  Recently, Melendez \& Ramirez\cite{mr} 
reanalyzed 62 halo dwarfs using an improved infrared flux method effective 
temperature scale.  While they failed to confirm the [Li] -- [Fe/H] 
correlation claimed by Ryan {\it et al.}\cite{ryan}, they suggest an even 
higher relic lithium abundance: [Li]$_{\rm P} = 2.37\pm0.05$.  A very
detailed and careful reanalysis of extant observations with great
attention to systematic uncertainties and the error budget has been
done by Charbonnel and Primas\cite{cp}, who find no convincing evidence 
for a Li trend with metallicity, deriving [Li]$_{\rm P} = 2.21\pm0.09$ 
for their full sample and [Li]$_{\rm P} = 2.18\pm0.07$ when they restrict
their sample to unevolved (dwarf) stars.  They suggest the Melendez \& 
Ramirez value should be corrected downwards by 0.08 dex to account for
different stellar atmosphere models, bringing it into closer agreement
with their results.  To err on the side of conservatism, the lithium
abundance of Melendez \& Ramirez\cite{mr}, [Li]$_{\rm P} = 2.37\pm0.05$,
which is closer to the SBBN expectation, will be adopted in further
comparisons.

There is tension between the SBBN predicted relic abundance of \7li 
([Li]$_{\rm P} = 2.65^{+0.05}_{-0.06}$; see \S\ref{bbnforped}) and that
derived from recent observational data ([Li]$_{\rm P} = 2.37\pm0.05$).  
Systematic errors may play a large role confirming or resolving this 
factor of two discrepancy.  The role of the stellar temperature scale 
has already been mentioned.  Another concern is associated with the 
temperature structures of the atmospheres of these very cool, metal-poor 
stars.  This can be important because a large ionization correction 
is needed since the observed neutral lithium is a minor component of 
the total lithium.  Furthermore, since the low metallicity, dwarf, 
halo stars used to constrain primordial lithium are among the oldest 
in the Galaxy, they have had the most time to alter (by dilution 
and/or destruction) their surface lithium abundances, as is seen 
to be important for many of the higher metallicity stars shown in 
Fig.~\ref{fig:livsfe}.  While mixing stellar surface material to the 
interior would destroy or dilute any prestellar lithium, the very 
small observed dispersion among the lithium abundances in the low 
metallicity halo stars (in contrast to the very large spread for the 
higher metallicity stars) suggests this correction may not be large 
enough ($\la 0.1-0.2$ dex at most) to bridge the gap between theory 
and observation; see, \eg Pinsonneault {\it et al.}\cite{pinsono} and 
further references therein.

\section{Discussion}

The cosmic nuclear reactor was active for a brief epoch in the early
evolution of the universe.  As the Universe expanded and cooled the
nuclear reactor shut down after $\sim$~20 minutes, having synthesized
in astrophysically interesting abundances only the lightest nuclides 
D, \3he, \4he, and \7li.  For the standard models of cosmology and 
particle physics (SBBN) the relic abundances of these nuclides depend 
on only one adjustable parameter, the baryon abundance parameter 
$\eta_{\rm B}$ (the post-\epm annihilation value of the baryon 
(nucleon) to photon ratio).  If the standard models are the 
correct description of the physics controlling the evolution 
of the universe, the abundances of the four nuclides should be 
consistent with a single value of $\eta_{\rm B}$ and this baryon 
density parameter should also be consistent with the values 
inferred from the later evolution of the universe (\eg at present 
as well as $\sim$~400 kyr after BBN, when the relic photons left 
their imprint on the CBR observed by WMAP and other detectors).  
There are, however, two other particle physics related cosmological 
parameters, the lepton asymmetry parameter \xie and the expansion 
rate parameter $S$, which can affect the BBN-predicted relic 
abundances.  For SBBN it is assumed that \xie = 0 and $S = 1$.  
Deviations of either or both of these parameters from their 
standard model expected values could signal new physics beyond 
the standard model(s).

The simplest strategy is to test first the predictions of SBBN.
Agreement between theory and observations would provide support 
for the standard models.  Disagreements are more difficult to 
interpret in that while they may be opening a window on new physics,
they may well be due to unaccounted for systematic errors along 
the path from observations of post-BBN material to the inferred 
primordial abundances.  Subject to this latter caveat, the 
confrontation between theory and data can provide useful limits 
to (some of) the parameters associated with new physics which 
complement those from high precision, terrestrial experiments. 
In the comparisons presented below, the abundances (and their
inferred uncertainties) presented in \S\ref{abund} are adopted and 
compared to the BBN predictions described by the simple fits from 
\S\ref{bbnforped}.  For \4he the SSBN range in $\eta_{\rm B}$ favored 
by the adopted primordial abundance lies outside the range of validity 
of the simple fit; for \4he  and SBBN, the best fit and uncertainty 
in $\eta_{\rm B}$ is derived from the more detailed BBN calculations.  
While not all models of new physics proposed in the literature can 
be tested in this manner, this approach does offer the possibility 
of constraining a large subset of them and of providing a useful 
framework for understanding qualitatively how many of the others 
might affect the BBN predictions.

\subsection{SBBN}

\begin{figure}
\centerline{\psfig{file=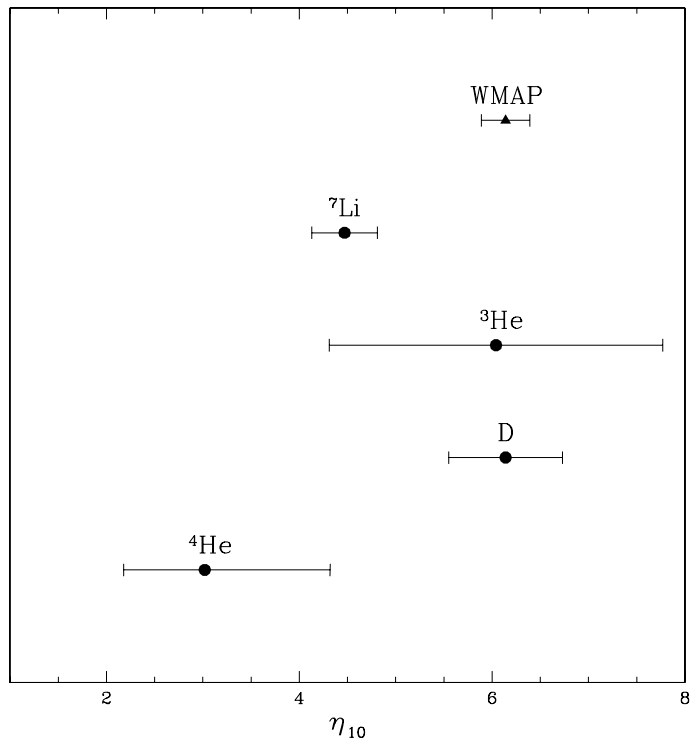,width=10.7cm}}%12cm}}
\vspace*{0pt}%\vspace*{8pt}
\caption{The SBBN values for the early universe ($\sim$~20 minutes) 
baryon abundance parameter $\eta_{10}$ inferred from the adopted 
primordial abundances of D, \3he, \4he, and \7li (see \S3.1-3.4).  
Also shown is the WMAP-derived CBR and LSS value ($\sim$~400 kyr).}
\label{fig:eta}  
\end{figure}    
 
The discussion in \S\ref{abund} identifies a set of primordial 
abundances.  Since these choices are certainly subjective and
likely to change as more data are acquired, along with a better
understanding of and accounting for systematic errors, the analytic
fits presented in \S\ref{bbnforped} can be very useful in relating
new conclusions and constraints to those presented here.  The
abundances and nominal $1 \sigma$ uncertainties adopted here are: 
\yd = $2.6 \pm 0.4$, $y_{3} = 1.1 \pm 0.2$, \Yp = $0.241 \pm 0.004$, 
and \yli = $2.34^{+0.29}_{-0.25}$ ([Li]$_{\rm P} = 2.37 \pm 0.05$).  
\Yp $\leq 0.254$ is adopted for an upper bound (at $\sim 2\sigma$) to 
the primordial \4he mass fraction.  The corresponding SBBN values 
of the baryon density parameter are shown in Figure~\ref{fig:eta}, 
along with that inferred from the CBR and observations of Large 
Scale Structure\cite{sperg} (labelled WMAP).

As Figure~\ref{fig:eta} reveals, the adopted relic abundances of D and 
\3he are consistent with the SBBN predictions ($\eta_{\rm D} = 6.1 \pm 
0.6$, $\eta_{^{3}{\rm He}} = 6.0 \pm 1.7$) and both are in excellent 
agreement with the non-BBN value\cite{sperg} ($\eta_{\rm WMAP} = 6.14 \pm 
0.25$).  If the most recent deuterium abundance determination in a high 
redshift, low metallicty QSOALS\cite{webb} is included in estimating the 
relic D abundance, the mean shifts to a slightly lower value (\yd = $2.4 
\pm 0.4$), corresponding to a slightly higher estimate for the baryon 
density parameter ($\eta_{\rm D} = 6.4 \pm 0.7$), which is still 
consistent with \3he and with WMAP.  Were it not for the very large 
dispersion among the D abundance determinations (see \S3.1), the formal 
error in the mean ($\sim 5$\%) could have been adopted for the uncertainty 
in $y_{\rm D}$, leading to a $\sim 3$\% determination of $\eta_{\rm D}$, 
competitive with that from WMAP.  Due to the very large observational 
and evolutionary uncertainties associated with \3he, its abundance 
mainly provides a consistency check at present.  Since the variations 
of its predicted relic abundance with $S$ and \xie are similar to those 
for D, \3he will not add new information to that from D in the comparisons 
to be discussed below.  

In addition to the successes of D and \3he, Figure~\ref{fig:eta} exposes 
a tension between WMAP (and D and \3he) and the adopted primordial 
abundances of \4he and \7li.  The $1\sigma$ range determined from \4he 
is low: $2.2 \leq \eta_{\rm He} \leq 4.3$; however, the $2\sigma$ range 
is much larger: $1.7 \leq \eta_{\rm He} \leq 6.4$, encompassing the 
WMAP-inferred baryon density.  The \7li inferred baryon density is also 
low ($\eta_{\rm Li} = 4.5 \pm 0.3$) and here the adopted errors appear 
to be far too small to bridge the gap to D and WMAP.  These tensions may 
be a sign of systematic errors introduced when the observational data is 
used to derive the inferred primordial abundances or, it could be a signal 
of new physics beyond the standard models of cosmology and particle physics.

\subsection{Lithium}

As identified above, the SBBN abundances of D and \3he are in agreement
with each other and with the non-BBN estimate of the baryon density
parameter from Large Scale Structure and the CBR.  However, while the
inferred primordial abundance of \4he is less than 2$\sigma$ away from
the SBBN-predicted value, that of lithium differs from expectations by
a factor of $\sim$~2 (or more).  It is unlikely that this 
conflict can be resolved through a non-standard expansion rate ($S \neq 
1$) or a non-zero lepton number ($\xi_{e} \neq 0$).  The reason is that 
in the $S - \eta_{\rm B}$ and $\xi_{e} - \eta_{\rm B}$ planes the 
isoabundance curves for D and \7li are very nearly parallel (see eqs.~16 
\& 23 in \S2.2 and Figs.~1 \& 2 from Kneller \& Steigman\cite{ks}), so that 
once \yd is constrained, there is very little freedom to modify $y_{\rm Li}$.  
This may be seen by combining eqs.~16 \& 23 to relate $\eta_{\rm Li}$ 
to $\eta_{\rm D}$,
\be
\eta_{\rm Li} = \eta_{\rm D} + 3[(S-1) - \xi_{e}].
\ee
Thus, for $\eta_{\rm D}~\ga 6$ and $|S-1|~\la 0.1$, $|\xi_{e}|~\la 
0.1$, $\eta_{\rm Li} \approx \eta_{\rm D}~\ga 6$, so that \yli $\ga 
4$ ([Li]$_{\rm P} ~\ga 2.6$).

Nonetheless, a non-standard physics explanation of the lithium conflict
is not ruled out.  Indeed, there are models where late-decaying, massive
particles reinitiate BBN, modifying the abundances of the light nuclides
produced during the first 20 minutes.  For an extensive, yet likely 
incomplete list (with apologies) of references, see Ref.~[36] and further 
references therein.  In such models it is quite possible to {\it reduce}
the original BBN abundance of \7li to bring it into agreement with the
value inferred from the observational data\cite{ryan,thor,bonif,mr}.
However, it is found that when the many new parameters available to these 
models are adjusted to achieve this agreement, the modified relic abundance
of \3he is much too large (see, \eg Ellis, Olive, and Vangioni\cite{feng}).

The difficulty in reconciling the observed and predicted relic abundances
of \7li suggests that the problem may be in the stars.  It is not at all
unexpected that the very old halo stars where lithium is observed will
have modified their original surface abundances, \7li in particular (see 
Pinsonneault {\it et al.}\cite{pinsono} and Charbonnel and Primas\cite{cp} for
discussions and many additional references).  While there is no dearth of
physical mechanisms capable of destroying or diluting surface lithium, 
many of which are supported by independent observational data, the challenge
has been to account for the required depletion (factor of 2 -- 3) while
maintaining a negligible dispersion ($\la 0.1$ dex) among the ``Spite 
plateau" lithium abundances.  

Another possibility for reconciling the observed and predicted relic 
abundances of \7li lies in the nuclear physics.  After all, given the
estimates of uncertainties in the cross sections of the key nuclear
reactions leading to the production and destruction of mass-7, the
BBN-predicted abundance of \7li is the most uncertain ($\sim 10-20$\%)
of all the light nuclides.  Perhaps the conflict between theory and 
observation is the result of an error in the nuclear physics.  This
possibility was investigated by Cyburt, Fields, and Olive\cite{cfo} who
noted that some of the same nuclear reactions of importance in BBN, 
play a role in the standard solar model and are constrained by its 
success in accounting for the observed flux of solar neutrinos.  While
the uncertainty of a key nuclear reaction (\3he($\alpha, \gamma)^{7}$Be)
is large ($\sim 30$\%), it is far smaller than the factor of $\sim 3$
needed to reconcile the predicted and observationally inferred 
abundances\cite{cfo}.

Considering the current state of affairs (no successful resolution based
on new physics; possible reconciliation based on stellar astrophysics),
\7li is not used below where the adopted relic abundances of D and \4he 
are employed to set constraints on $S$ and/or $\xi_{e}$.

\subsection{Non-Standard Expansion Rate: $S \neq 1$ ($\xi_{e} = 0$)}
\label{s}

\begin{figure}
\centerline{\psfig{file=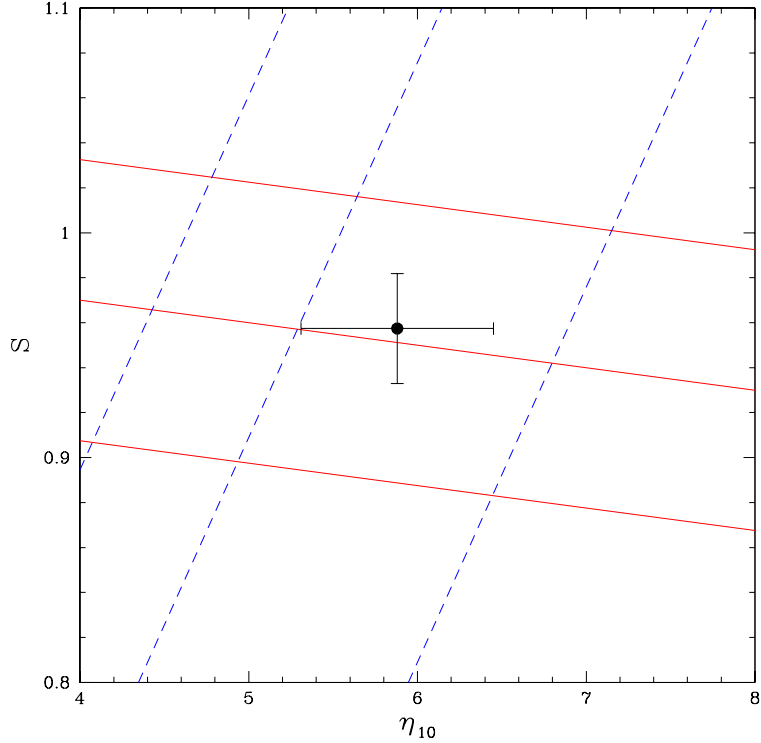,width=10.7cm}}%12cm}}
\vspace*{0pt}%\vspace*{8pt}
\caption{The D and \4he isoabundance curves in the $S - \eta_{10}$ 
plane, as in Fig.~\ref{fig:svseta}.  The best fit point and the
error bars correspond to the adopted primordial abundances of D 
and \4he.}
\label{fig:svsetadhe}  
\end{figure}    

If the lepton asymmetry is very small, of order the baryon asymmetry, 
then BBN depends on only two free parameters, $\eta_{\rm B}$ and $S$ 
(or N$_{\nu}$).  Since the primordial abundance of D largely probes 
$\eta_{\rm B}$ while that of \4he is most sensitive to $S$ (see Fig.~2 
and eqs.~16 \& 20), for each pair of \yd and \Yp values (within reason) 
there will be a corresponding pair of $\eta_{\rm B}$ and $S$ values.  
For the D and \4he abundances adopted above (\yd $= 2.6\pm0.4$, \Yp 
$= 0.241\pm0.004$) the best fits for $\eta_{\rm B}$ and $S$, shown 
in Figure~\ref{fig:svsetadhe}, are for $\eta_{10} = 5.9\pm0.6$ and 
$S = 0.96\pm0.02$; the latter corresponds to N$_{\nu} = 2.5\pm0.3$.  
These values are completely consistent with those inferred from the 
joint constraints on $S$ and $\eta_{\rm B}$ from WMAP\cite{barg1}. 

As expected from the discussion in \S2.2, the lithium abundance is largely
driven by the adopted deuterium abundance and is little affected by the
small departure from the standard expansion rate.  For the above best fit 
values, \yli $= 4.3\pm0.9$ ([Li]$_{\rm P} = 2.63^{+0.08}_{-0.10}$).  This 
class of non-standard models ($S \neq 1$), while reconciling \4he with
D and with the CBR, is incapable of resolving the lithium conflict.

\subsection{Non-Zero Lepton Number: $\xi_{e} \neq 0$ ($S = 1$)}
\label{xi}

\begin{figure}
\centerline{\psfig{file=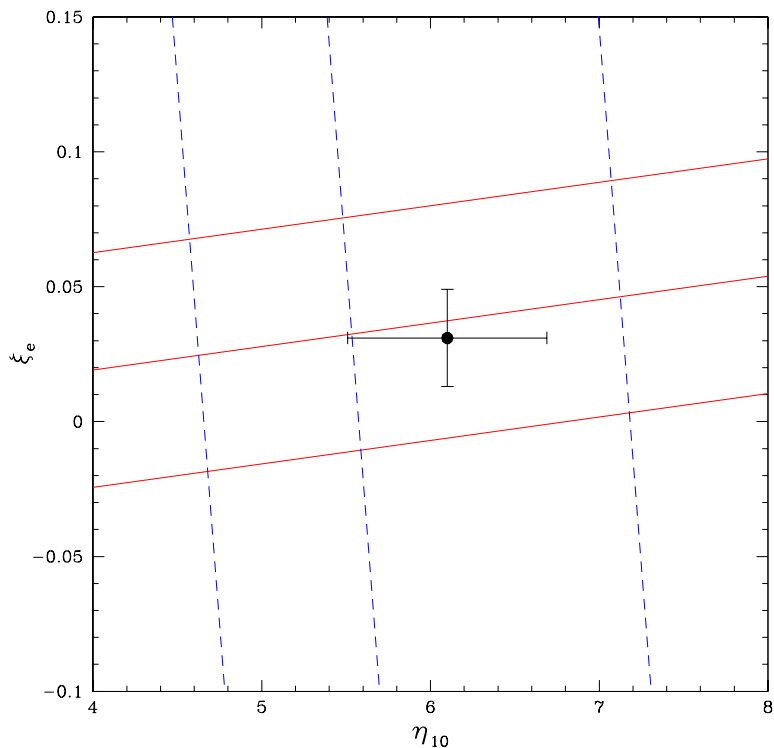,width=10.7cm}}%12cm}}
\vspace*{0pt}%\vspace*{8pt}
\caption{The D and \4he isoabundance curves in the $\xi_{e} - \eta_{10}$ 
plane, as in Fig.~\ref{fig:xivseta}. The best fit point and the error 
bars correspond to the adopted primordial abundances of D and \4he.}
\label{fig:xivsetadhe}  
\end{figure}    

While most popular extensions of the standard model which attempt to 
account for neutrino masses and mixings suggest a universal lepton
asymmetry comparable in magnitude to the baryon asymmetry ($\xi_{e} 
\sim O(\eta_{\rm B}) ~\la 10^{-9}$)\footnote{By charge neutrality the 
charged lepton excess is equal to the proton excess which constitutes 
$\ga 87$\% of the baryon excess.  Therefore, any significant lepton 
asymmetry ($\xi_{e} >> \eta_{\rm B}$) must be hidden in the unobserved 
relic neutrinos.}, there is no direct evidence that nature has made 
this choice.  Although the CBR is blind to a relatively small lepton 
asymmetry, BBN provides an indirect probe of it\cite{barg2}.  As discussed 
in \S1.3 \& \S2, a lepton asymmetry can change the neutron to proton 
ratio at BBN, modifying the light element yields, especially that of 
\4he.  Assuming $S=1$ and allowing $\xi_{e} \neq 1$, BBN now depends 
on the two adjustable parameters $\eta_{\rm B}$ and $\xi_{e}$ which 
may be constrained by the primordial abundances of D and \4he.  Given 
the strong dependence of \Yp on $\xi_{e}$ and of \yd on $\eta_{\rm B}$, 
these nuclides offer the most leverage.  In Figure~\ref{fig:xivsetadhe} 
are shown the D and \4he isoabundance curves (\ie the 
fits from \S2.2) in the $\xi_{e} - \eta_{\rm B}$ plane, along with 
the best fit point (and its $1\sigma$ uncertainties) determined by 
the adopted primordial abundances.  The best fit baryon abundance, 
$\eta_{10} = 6.1\pm0.6$ is virtually identical to the SBBN (and WMAP) 
value.  While the best fit lepton asymmetry, $\xi_{e} = 0.031\pm0.018$, 
is non-zero, it differs from zero by less than $2\sigma$ (as it should 
since the adopted value of \Yp differs from the SBBN expected value by 
less than $2\sigma$).

As expected from the discussion above for $S \neq 1$ and in \S2.2, here, 
too, the lithium abundance is largely driven by the adopted deuterium 
abundance and is little affected by the small lepton asymmetry allowed
by D and \4he.  For the above best fit values, the predicted lithium 
abundance is virtually identical to the SBBN/WMAP and $S \neq 1$ values: 
\yli $= 4.3\pm0.9$ ([Li]$_{\rm P} = 2.64^{+0.08}_{-0.10}$).  A lepton
asymmetry which reconciles \4he with D cannot resolve the lithium conflict.

\subsection{An Example: Alternate Relic Abundances for D and \4he}

It is highly likely that at least some of the tension between D and 
\4he is due to errors associated with inferring their primordial 
abundances from the current observational data.  As a result, in 
the future the abundances adopted here may be replaced by revised
estimates.  This is where the simple, analytic fits derived by 
KS\cite{ks} and presented in \S\ref{bbnforped} can be of value to those 
who lack an in-house BBN code.  Provided that the {\it revised} 
abundances lie in the ranges $2~\la y_{\rm D}~\la 4$ and $0.23~\la 
$Y$_{\rm P}~\la 0.25$, these fits will provide quite accurate, back 
of the envelope estimates of $\eta_{\rm B}$, $S$, $\xi_{e}$, and of 
$y_{\rm Li}$.  As an illustration, let's revisit the discussion in 
\S\ref{s} \& \S\ref{xi}, now adopting for D the weighted mean deuterium 
abundance which results when the most recent determination\cite{webb} 
is included, \yd $= 2.4\pm0.4$ (see \S\ref{d}), along with, for \4he, 
the helium abundance derived by applying the OS mean offset to the 
IT-inferred primordial value (see eq.~25) {\bf without} the {\it 
icf}-correction, \Yp $= 0.2472\pm0.0035$ (see \S\ref{4he}).  These 
alternate abundances correspond to $\eta_{\rm D} \approx 6.5\pm0.7$ 
and $\eta_{\rm He} \approx 5.5\pm2.2$.

For \xie = 0, the {\it new} values for the expansion rate factor 
and the baryon density parameter are $S = 0.991\pm0.022$ (N$_{\nu} 
= 2.9\pm0.3$) and $\eta_{10} = 6.4\pm0.6$.  While the \Nnu
estimate is entirely consistent with \Nnu = 3, the corresponding 
$\sim 2\sigma$ upper bound (N$_{\nu}~\la 3.5$) still excludes 
even one extra light scalar.  The baryon density parameter is 
slightly higher than, but entirely consistent with that inferred 
from the CBR.  As anticipated from the previous discussion, the 
predicted lithium abundance hardly changes at all, but it does 
increase slightly to further exacerbate the conflict with the 
observationally inferred value, \yli $\approx 4.9\pm1.1$ 
([Li]$_{\rm P} = 2.69^{+0.09}_{-0.11}$).

For $S$ = 0, the {\it new} values for the lepton asymmetry parameter 
and the baryon density parameter are $\xi_{e} = 0.007\pm0.016$ and 
$\eta_{10} = 6.5\pm0.7$.  The former is consistent with no lepton 
asymmetry (\ie with $\xi_{e} \sim O(\eta_{\rm B})$) and the latter 
is slightly higher than, but still entirely consistent with the
baryon density parameter inferred from the CBR.  As expected, 
here, too, the predicted lithium abundance increases slightly
from the already too large SBBN value, \yli $\approx 4.9\pm1.1$ 
([Li]$_{\rm P} = 2.69^{+0.09}_{-0.11}$).

\subsection{Other Non-Standard Models}

Although the parameterization of BBN in terms of $S$ and \xie explored in 
the previous sections encompasses a large set of non-standard models of
cosmology and particle physics, it by no means describes all interesting
extensions of the standard model.  As already mentioned, there is a class
of models where BBN proceeds normally but a second epoch of early universe 
nucleosynthesis is initiated by the late decay of a massive particle\cite{feng}.  
Despite the fact that such models have many more free parameters, such as 
the mass, abundance, and lifetime of the decaying particle, the constraints 
imposed by the observationally inferred relic abundances of D, \3he, \4he, 
and \7li are sufficiently strong to challenge them (see, \eg Ellis, Olive 
\& Vangioni\cite{feng}).  

There are other models which cannot be simply described by the \{$\eta_{\rm 
B}$, $S$, $\xi_{e}$\} parameter set.  In most cases they introduce several free 
parameters in addition to the baryon density parameter.  Since there are 
only four nuclides whose relic abundances are reasonably constrained, the 
leverage of BBN on these models may be limited in some cases.  A case in 
point is the class of models where the universe is inhomogeneous at BBN 
(IBBN); see the recent article by Lara\cite{lara} and the extensive 
references to earlier work therein.   In IBBN the geometry of the 
inhomogeneities (spheres, cylinders, ...) is important, as are the 
scales of the imhomogeneities and their amplitudes (density contrasts).  
Nonetheless, even with all these adjustable parameters, except when 
they take on values indistinguishable from SBBN, IBBN models predict 
an excess of lithium (even more of an excess than for SBBN).  This is 
inevitable since in IBBN \7li is overproduced in the low nucleon density 
regions and $^{7}$Be is overproduced in the high density regions (see 
the multi-valued lithium abundance curve in Fig.~\ref{fig:schrplot}).

\section{Summary and Conclusions}

As the early Universe evolved from hot and dense to cool and dilute,
it passed through a short-lived epoch when conditions of temperature
and density permitted the synthesis of astrophysically interesting 
abundances of D, \3he, \4he, and \7li.  At present, observations of 
these nuclides in a variety of astrophysical sites (stars, Galactic 
and extragalactic \hii regions, QSOALS, etc.) have permitted quite 
precise estimates of their primordial abundances, opening a window 
on early-universe cosmology and providing constraints on physics 
beyond the standard models of cosmology and particle physics.  The 
relic abundances of D, \3he, and \7li are nuclear reaction rate
limited, providing good probes of the nucleon density when the 
universe was less than a half hour old.  This is complementary to
baryon density estimates from the CBR, whose information was encoded
some 400 kyr later, and to those provided by observations of the current
or recent universe, 13 Gyr after BBN.  Of these light nuclides, D is
the baryometer of choice and current estimates of the relic abundance,
\yd $=2.6\pm0.4$, suggest a baryon to photon ratio, unchanged from BBN
to the present, of $\eta_{10} = 6.1\pm0.6$, in excellent agreement with
the WMAP and LSS determined value of $\eta_{10} = 6.14\pm0.25$.  For 
this choice ($S=1$, \xie = 0), the less well constrained relic abundance 
of \3he ($y_{3} = 1.1\pm0.2$)\cite{brb} is also in agreement with its SBBN
expected value.  These successes for the standard models are tempered by
the challenges posed by \4he and \7li.  The SBBN predicted \4he mass
fraction, \Yp $=0.248$, differs from the observationally inferred primordial
abundance adopted here, \Yp $= 0.241\pm0.004$, by nearly $2\sigma$.  However, 
as discussed in \S\ref{4he}, the uncertainties in \Yp are likely dominated 
by systematic, not statistical errors, so it is difficult to know if this 
tension between D (and \3he and the CBR) and \4he is cause for serious 
concern.  In contrast to the other light nuclides, the BBN abundance of
\4he is insensitive to the nuclear reaction rates and, hence, to the
nucleon density at BBN.  \Yp is largely set by the neutron to proton
ratio at BBN, so that \4he is an excellent probe of the weak interactions
and of the early Universe expansion rate.  Perhaps the \4he challenge to
SBBN is a signal of new physics.  The SBBN predicted abundance of \7li 
([Li]$_{\rm P} = 2.65^{+0.05}_{-0.06}$) is nearly a factor of two higher 
than the observationally determined value\cite{mr} ([Li]$_{\rm P} = 
2.37\pm0.05$).  While there is some spread in the lithium abundances 
inferred from the data\cite{cp}, the largest cause for concern is that 
Li is observed in the oldest stars in the Galaxy, which have had ample 
time to modify their original surfaces abundances.  For \7li, it appears 
likely that astrophysical uncertainties dominate at present.  

Thus, while there appears to be qualitative confirmation of the standard 
models of cosmology and particle physics extrapolated back to the first 
seconds of the evolution of the Universe, precision should not be confused 
with accuracy.  The accuracy of the presently-inferred primordial abundances
of D, \3he, \4he, and \7li remains in question and it would not be at all
surprising if one or more of them changed by more than the presently-quoted
errors.  After all, there are only 5 (6) lines of sight where deuterium 
is observed in high-redshift, relatively unprocessed (low metallicity) 
material; \3he is only observed in the chemically processed interstellar 
medium of the Galaxy and the lack of variation of its abundance with 
metallicity or with position in the Galaxy suggest a very delicate balance 
between post-BBN production, destruction and survival; systematic errors 
and corrections to the  \4he abundance inferred from observations of low 
metallicity, extragalactic \hii regions are likely larger, maybe much 
larger, than the current statistical uncertainties; lithium is derived from 
observations of very old, very low metallicity stars (good!) in our Galaxy 
(bad?) and the corrections for stellar atmosphere models and, especially, 
for main sequence mixing-induced depletion and destruction remain large 
and uncertain.  Much interesting, important work remains for observational
and theoretical astronomers. 

The current standard models receive strong support from
these messengers from the early universe, confirming in broad outline 
our understanding of the evolution of the Universe and the particles 
in it, from the first seconds to the present.  Any models of new physics
must consider this success and avoid introducing new conflicts.  Much
interesting, important work remains for cosmologists and high energy
theorists.

\section*{Acknowledgments}

I am grateful to all those I've collaborated with over the years on 
the subject of this review.  Discussions with J.~E.~Felten, K.~A.~Olive, 
E.~D.~Skillman, M.~Tosi, D.~Tytler, and J.~K.~Webb were especially 
valuable in its preparation. My research is supported by 
%the DOE.
a grant (DE-FG02-91ER40690) from the US Department of Energy.

\end{document}